\documentclass[aps, 10pt, twocolumn, showpacs,preprintnumbers,amsmath,amssymb,APSl,prd,nofootinbib,superscriptaddress]{revtex4-2}

\usepackage{graphicx,color}
\usepackage{subfig}
\usepackage{amssymb}
\usepackage{hyperref}
\usepackage[utf8]{inputenc}
\usepackage[english]{babel}
\usepackage{epsfig}
\usepackage{wasysym}
\usepackage{color,xcolor}
\usepackage{amsmath}
\usepackage{bm}
\usepackage{epsfig}
\usepackage{amsfonts}
\usepackage{dcolumn}
\usepackage{float}

\hypersetup{colorlinks=true, linkcolor=blue, citecolor=green}

\begin{document}

\title{Novel charged black hole solutions in conformal Killing gravity\\}

	\author{Ednaldo L. B. Junior} \email{ednaldobarrosjr@gmail.com}
\affiliation{Faculdade de F\'{i}sica, Universidade Federal do Pará, Campus Universitário de Tucuruí, CEP: 68464-000, Tucuruí, Pará, Brazil}

     \author{José Tarciso S. S. Junior}
    \email{tarcisojunior17@gmail.com}
\affiliation{Faculdade de F\'{i}sica, Programa de P\'{o}s-Gradua\c{c}\~{a}o em F\'{i}sica, Universidade Federal do Par\'{a}, 66075-110, Bel\'{e}m, Par\'{a}, Brazill}

	\author{Francisco S. N. Lobo} \email{fslobo@ciencias.ulisboa.pt}
\affiliation{Instituto de Astrof\'{i}sica e Ci\^{e}ncias do Espa\c{c}o, Faculdade de Ci\^{e}ncias da Universidade de Lisboa, Edifício C8, Campo Grande, P-1749-016 Lisbon, Portugal}
\affiliation{Departamento de F\'{i}sica, Faculdade de Ci\^{e}ncias da Universidade de Lisboa, Edif\'{i}cio C8, Campo Grande, P-1749-016 Lisbon, Portugal}

    \author{\\Manuel E. Rodrigues} \email{esialg@gmail.com}
\affiliation{Faculdade de F\'{i}sica, Programa de P\'{o}s-Gradua\c{c}\~{a}o em F\'{i}sica, Universidade Federal do Par\'{a}, 66075-110, Bel\'{e}m, Par\'{a}, Brazill}
\affiliation{Faculdade de Ci\^{e}ncias Exatas e Tecnologia, Universidade Federal do Par\'{a}, Campus Universit\'{a}rio de Abaetetuba, 68440-000, Abaetetuba, Par\'{a}, Brazil}

     \author{Luís F. Dias da Silva} 
        \email{fc53497@alunos.fc.ul.pt}
\affiliation{Instituto de Astrof\'{i}sica e Ci\^{e}ncias do Espa\c{c}o, Faculdade de Ci\^{e}ncias da Universidade de Lisboa, Edifício C8, Campo Grande, P-1749-016 Lisbon, Portugal}
    
    \author{Henrique A. Vieira} \email{henriquefisica2017@gmail.com}
\affiliation{Faculdade de F\'{i}sica, Programa de P\'{o}s-Gradua\c{c}\~{a}o em F\'{i}sica, Universidade Federal do Par\'{a}, 66075-110, Bel\'{e}m, Par\'{a}, Brazill}


\begin{abstract}
In this paper, we investigate static spherically symmetric solutions in the context of Conformal Killing Gravity, a recently proposed modified theory of gravity that offers a new approach to the cosmological constant problem. Coupling this new theory with nonlinear electrodynamics, we derive the corresponding field equations and study their behavior under different parameter choices. We analyze three different models, each focusing on different key parameters.
Our results reveal a rich causal structure with multiple horizons and transitions between extreme and non-extreme solutions depending on the parameter values. Moreover, we compute the nonlinear Lagrangian density for each model and find that it agrees with Maxwell theory in the limit $F \rightarrow 0$. We also confirm the existence of a central curvature singularity via the Kretschmann scalar. To connect our theoretical results with observational prospects, we study the black hole shadows associated with each model. The analysis shows that the calculated shadow size and shape of the three proposed models are consistent with the data for the supermassive object at the center of our galaxy and are therefore possible candidates for modeling this structure.
\end{abstract}

\date{\today}

\maketitle

\section{Introduction}

The General Theory of Relativity (GR) has proved remarkably successful in explaining gravitational phenomena at various scales. It has passed several tests over the last century \cite{Will:2018mcj,Will:2014zpa}, 
and has recently received significant experimental evidence with the detection of the first signals of gravitational waves by the LIGO and VIRGO collaborations \cite{LIGOScientific:2016aoc,LIGOScientific:2017vwq} and the very first image of superheated plasma swirling around the supermassive object at the core of the galaxy M87 \cite{fotoBN1,fotoBN2,fotoBN3,fotoBN4,fotoBN5,fotoBN6,fotoBN7,fotoBN8,fotoBN9} and of our own galaxy \cite{fotosrg1,fotosrg2,fotosrg3,fotosrg4,fotosrg5,fotosrg6}. 
However, GR still encounters difficulties when trying to reconcile it with quantum mechanics and fully explain dark energy and dark matter, concepts created to explain the experimental data pointing to an accelerating expanding universe \cite{Perlmutter, Riess, Garnavich} and the behavior of the galactic rotation curves
and the mass discrepancy in clusters of galaxies, respectively. There is no theoretical possibility that has been widely accepted as an explanation for this phenomenon. Therefore, significant efforts have been made in recent years to modify this theory to address the challenges that arise in cosmological and high-energy domains \cite{Sotiriou:2008rp,Clifton:2011jh}. Indeed, modifications of GR offer an alternative framework that could provide insights into these open problems.

One of these attempts is the so-called ``Cotton Gravity"' (CG)  proposed by Harada~\cite{Harada:2021bte}. In this formulation, the effects describing gravity are attributed to the Cotton tensor, and the field equations contain third-order derivatives. 
This higher-order nature introduces distinct features compared to standard GR. Interestingly, any solution of the Einstein equations, with or without the cosmological constant, also satisfies the field equations described by the Cotton tensor within the framework of this theory. In this context, the cosmological constant is reinterpreted as an integration constant, providing a new perspective on its role in the theory.
Similar to the Schwarzschild solution in GR, Harada also found the first non-trivial solution of CG with spherical and static symmetry.
Building on this work, Harada further explored the implications of CG in subsequent research \cite{Harada:2022edl}. Notably, he set aside the consideration of dark matter and applied the post-Newtonian limit of CG to numerically solve the field equations. This approach was specifically aimed at interpreting the observed rotation curves of several galaxies, providing a potential alternative explanation to the conventional dark matter paradigm.

Another promising line of research is the recently proposed Conformal Killing Gravity (CKG) \cite{Harada,Barnes:2023uru,Mantica:2023stl}, a modified theory of gravity that provides additional flexibility in modeling cosmological and astrophysical phenomena. A key feature of CKG is the ability to treat the cosmological constant as an integration constant rather than assuming it as a fixed parameter. This feature provides a natural interpretation of dark energy and opens up new possibilities for solving the problem of the cosmological constant. 
Recently, nonlinear electrodynamics (NED) and scalar fields were used to construct models of regular black holes \cite{Junior:2023ixh} and black bounces \cite{Junior:2024vrv} within the framework of CKG theory. These developments represent significant progress in addressing the fundamental issues of singularities in gravitational physics.

Nonlinear electrodynamics, originally introduced as a modification of classical Maxwell's electrodynamics, was proposed by Born and Infeld in 1934 \cite{Born:1933pep,Born:1934gh}. This approach aimed to resolve the singularities inherent in Maxwell's theory, such as those associated with the central point charge and the self-energy of charges, by introducing a finite upper bound on the field strength. Following this foundational work, numerous other important contributions further advanced the field. For example, the Euler-Heisenberg formulation, driven by insights from quantum field theory, explored the quantum corrections to electrodynamics in the presence of strong fields \cite{Heisenberg:1936nmg}. Similarly, Plebanski’s work provided a deeper mathematical framework for nonlinear extensions of electrodynamics \cite{Plebanski}. These efforts were complemented by various generalizations and modern developments in NED, such as those by Kruglov and Bandos, which have enriched our understanding of its implications for both classical and quantum physics \cite{Kruglov:2014hpa,Kruglov:2014iwa,Kruglov:2014iqa,Bandos:2020jsw}.

More recently, high-frequency Very Long Baseline Interferometry (VLBI) observations of the supermassive black holes M87* \cite{EventHorizonTelescope:2019dse} and Sgr $A^{\star}$ \cite{EventHorizonTelescope:2022wkp,EventHorizonTelescope:2022xqj} have opened an unprecedented avenue for exploring and rigorously testing novel aspects of gravitational physics. These groundbreaking images, obtained through the collaborative efforts of the Event Horizon Telescope (EHT) collaboration, provide an level of detail regarding the shadow and the surrounding emission of these astrophysical objects. These observations have facilitated a deeper understanding of the behavior of gravity in the strong-field regime, offering unique insights into the nature of spacetime near black hole horizons and providing a critical platform for examining the predictions of GR and alternative gravitational theories in extreme environments.

In this work, we aim to construct spherically symmetric solutions within the framework of CKG theory coupled to nonlinear electrodynamics (NED). Our primary goal is to use these solutions to constrain the parameters of this alternative gravitational geometry by comparing the predicted shadow sizes with the observational estimates provided by the EHT collaboration. The shadow, which serves as a powerful probe of the spacetime geometry near the event horizon, is particularly sensitive to deviations from general relativity and provides a stringent test for alternative theories of gravity.
To achieve this, we adopt a methodology similar to that used in related works (e.g., \cite{Shaikh:2022ivr, Vagnozzi:2022moj, Ghosh:2022kit}), which leverage observational data on black hole shadows to derive constraints on theoretical models. By analyzing the shadow size and its dependence on the parameters of the CKG-NED solutions, we seek to place meaningful bounds on these parameters and assess the viability of this alternative gravitational framework in light of recent high-resolution observational data.

This work is organized as follows: In Sec. \ref{sec2}, we provide a concise introduction to the field equations of Conformal Killing gravity (CKG) coupled to nonlinear electrodynamics (NED). Additionally, we present a general methodology for deriving a Lagrangian within the context of this theory. In Sec. \ref{sec:bhsol}, we propose and thoroughly analyze solutions corresponding to black holes within this framework. In Sec. \ref{sec:rs}, we employ observational data from the Event Horizon Telescope (EHT) collaboration pertaining to Sgr $A^{\star}$ to impose constraints on the parameters characterizing these solutions. Finally, in Sec. \ref{sec:conclu}, we summarize our findings and present the conclusions of this study.

\section{CKG coupled to non-linear electrodynamics}\label{sec2}

\subsection{Field equations}

Recently, Harada proposed a novel modification of the theory of gravity known as Conformal Killing Gravity (CKG) \cite{Harada}, which was further explored in \cite{Mantica:2023stl}. This theory fulfills several critical theoretical criteria for gravitational models beyond GR, namely: (i) the cosmological constant emerges as an integration constant; (ii) the conservation of the energy-momentum tensor, $\nabla_\mu T^\mu_{\phantom{\mu}\alpha}=0$, follows directly from the gravitational field equation instead of being postulated; and (iii) a conformally flat metric is not necessarily a vacuum solution.

The field equations of CKG are described by 
\begin{equation} H_{\alpha\mu\nu} = 8\pi GT_{\alpha\mu\nu},\label{eq_CKG} \end{equation}
where $G$ is the gravitational constant and we use natural units in this work, i.e. $G=c=1$. The tensors $H_{\alpha\mu\nu}$ and $T_{\alpha\mu\nu}$ are defined as
\begin{eqnarray}
H_{\alpha\mu\nu}&\equiv & \nabla_{\alpha}R_{\mu\nu}+\nabla_{\mu}R_{\nu\alpha}+\nabla_{\nu}R_{\alpha\mu}
	\nonumber \\
&&-\frac{1}{3}\left(g_{\mu\nu}\partial_{\alpha}+g_{\nu\alpha}\partial_{\mu}+g_{\alpha\mu}\partial_{\nu}\right)R,
	 \\
T_{\alpha\mu\nu}&\equiv & \nabla_{\alpha}T_{\mu\nu}+\nabla_{\mu}T_{\nu\alpha}+\nabla_{\nu}T_{\alpha\mu}
	\nonumber \\
&&-\frac{1}{6}\left(g_{\mu\nu}\partial_{\alpha}+g_{\nu\alpha}\partial_{\mu}+g_{\alpha\mu}\partial_{\nu}\right) T \,,
\end{eqnarray}
where $R_{\mu\nu}$ and $T_{\mu\nu}$ represent the Ricci tensor and the energy-momentum tensor with their corresponding traces $R$ and $T$. In particular, $H_{\alpha\mu\nu}$ is totally symmetric in $\alpha$, $\mu$ and $\nu$ and satisfies $g^{\mu\nu}H_{\alpha\mu\nu}=0$ \cite{Harada}. Since $T_{\alpha\mu\nu}$ shares this symmetry, thus $g^{\mu\nu} T_{\alpha\mu\nu} =2 \nabla_\mu T^\mu_{\phantom{\mu}\alpha}$ obeys the conservation law $\nabla_\mu T^\mu_{\phantom{\mu}\alpha}=0$. In addition, the solutions of GR are also solutions in CKG \cite{Harada}.

Considering a static and spherically symmetric metric 
Harada derived the exact vacuum solution of the field equation $H_{\alpha\mu\nu} = 0$, expressed as \cite{Harada} 
\begin{equation} 
A(r)= 1 - \frac{2M}{r} - \frac{\Lambda}{3}r^2-\frac{\lambda}{5}r^4. 
\end{equation}
Here the term $2M/r$ corresponds to the Schwarzschild solution, while $\Lambda r^2/3$ denotes a de Sitter term, with the cosmological constant $\Lambda$ serving as the integration constant. The last term with $\lambda$ represents a new property of the theory and dominates at $r \rightarrow \infty$. In the case $\lambda = 0$, the solution reduces to the standard Schwarzschild-de Sitter solution from GR. In addition, the most general spherically symmetric static vacuum solution within this theory was derived in \cite{Barnes:2023uru}.

Recent work has also shown that Eq.~(\ref{eq_CKG}) is equivalent to Einstein's equation with any conformal Killing tensor, reducing the nature of the third-order equation to a second-order problem with respect to the metric tensor \cite{Mantica:2023stl}, which provides a simplified approach to finding solutions in Harada's theory.
As mentioned above, in this work we couple CKG to NED as a matter source applied to the energy-momentum tensor in the field equations~\eqref{eq_CKG}
 with
\begin{align}
    T{}_{\mu\nu}&=g_{\mu\nu}{\cal L}_{\rm NED}(F)-{\cal L}_{F}F_{\mu\alpha}F_{\nu}^{\phantom{\nu}\alpha},\\
T&=4{\cal L}_{\rm NED}(F)-4{\cal L}_{F}F\,.
\end{align}
\par
 ${\cal L}_{\rm NED}(F)$ is an arbitrary NED Lagrangian density that depends on the electromagnetic scalar $F=\frac{1}{4}F^{\mu\nu}F_{\mu\nu}$ and leads to Maxwell theory at $F \rightarrow 0$.
The antisymmetric Maxwell-Faraday tensor is defined by $F_{\mu \nu} = \partial_\mu A_\nu -\partial_\nu A_\mu$, where ${A_\alpha}$ is the electromagnetic vector potential.
We also introduce the following relevant expressions resulting from the influence of the gravitational field:
\begin{align}
\nabla_\mu ({\cal L}_F F^{\mu\nu})&=\frac{1}{\sqrt{-g} }\partial_\mu (\sqrt{-g} {\cal L}_F F^{\mu\nu})=0,\label{sol2}\\
     2\nabla_\nu\nabla^\mu \varphi&=-\frac{dV(\varphi)}{d\varphi}\,,
\end{align}
where ${\cal L}_F=\partial {\cal L} _{\rm NED}(F)/\partial F$.

In the solutions below, we will also consider the following useful consistency relationship
\begin{equation}
    {\cal L}_F-\frac{\partial {\cal L} _{\rm NED}}{\partial r} \bigg(\frac{\partial F}{\partial r}\bigg)^{-1}=0.\label{eq:RC}
\end{equation}

\par

\subsection{General solution for NED}

Our goal is to find a Lagrangian density that can generate new solutions in this theory.
Let us first consider the following static and spherically symmetric metric:
\begin{equation}
ds^2=A(r)dt^2-\frac{1}{A(r)}dr^2- C(r)\, d\Omega^2,\label{eq:ds}
\end{equation}
where the metric functions $A(r)$ and $C(r)$ depend only on the radial coordinate $r$ and $d\Omega^2\equiv d\theta^{2}+\sin^{2}\left(\theta\right)d\phi^{2}$.
Furthermore, we only consider solutions described by the magnetic charge, where the components for $F_{\mu\nu}$ and the electromagnetic scalar are $F_{23}=q \sin\theta$
and the electromagnetic scalar is given by
\begin{equation}
    F=\frac{q^2}{2 \Sigma^4(r)}.
\end{equation}

If we insert Eq. \eqref{eq:ds} with $C(r) =r^2$ into the Harada field equation \eqref{eq_CKG}, we obtain the following non-trivial equations
\begin{eqnarray}
		 r^5 A^{(3)}(r)-2 r \Big[r^2 \left(r \left(A''(r)+2 r {\cal L}'(r)\right)+A'(r)\right)
			\nonumber \\
		 +2 q^2
		{\cal L}_F'(r)+4 r \Big]+8 r^2 A(r)+16 q^2 {\cal L}_F(r) =0, 
		 \\
		 r^3 \left[r \left(r A^{(3)}(r)-2 A''(r)+2 r {\cal L}'(r)\right)-2 A'(r)\right]
		\nonumber \\ 
		+8 r^2 A(r)-4 r \left(q^2
		{\cal L}_F'(r)+2 r\right)+28 q^2 {\cal L}_F(r)=0.
	\label{eq:motion}
\end{eqnarray}

We can solve this system to find $\cal L$ and ${\cal L}_F$ without necessarily requiring that ${\cal L}_F=\partial {\cal L} _{\rm NED}(F)/\partial F$. The result is:
\begin{equation}
    \begin{aligned}
        & {\cal L} = f_0- \frac{r A'(r)+A(r)-1}{2 r^2}-f_1 q^2 r^2, \\
       & {\cal L}_F = \frac{r^2 \left(r^2 A''(r)-2 A(r)+2\right)}{4 q^2}+f_1 r^6.
    \end{aligned}
    \label{eq:lagrangianageral}
\end{equation}
We see that these functions fulfill the consistency conditions \eqref{eq:RC} and that we are free to choose the metric functions $A(r)$. The constant $f_0$ has the dimension $[L]^{-1}$ and $f_1$ has the dimension $[L]^{-3}$ in the system of geometrized units.

\section{ BLACK HOLE SOLUTIONS \label{sec:bhsol}}

\subsection{Model 1}

\begin{figure*}[t!]
	\includegraphics[width=7cm, height=6cm]{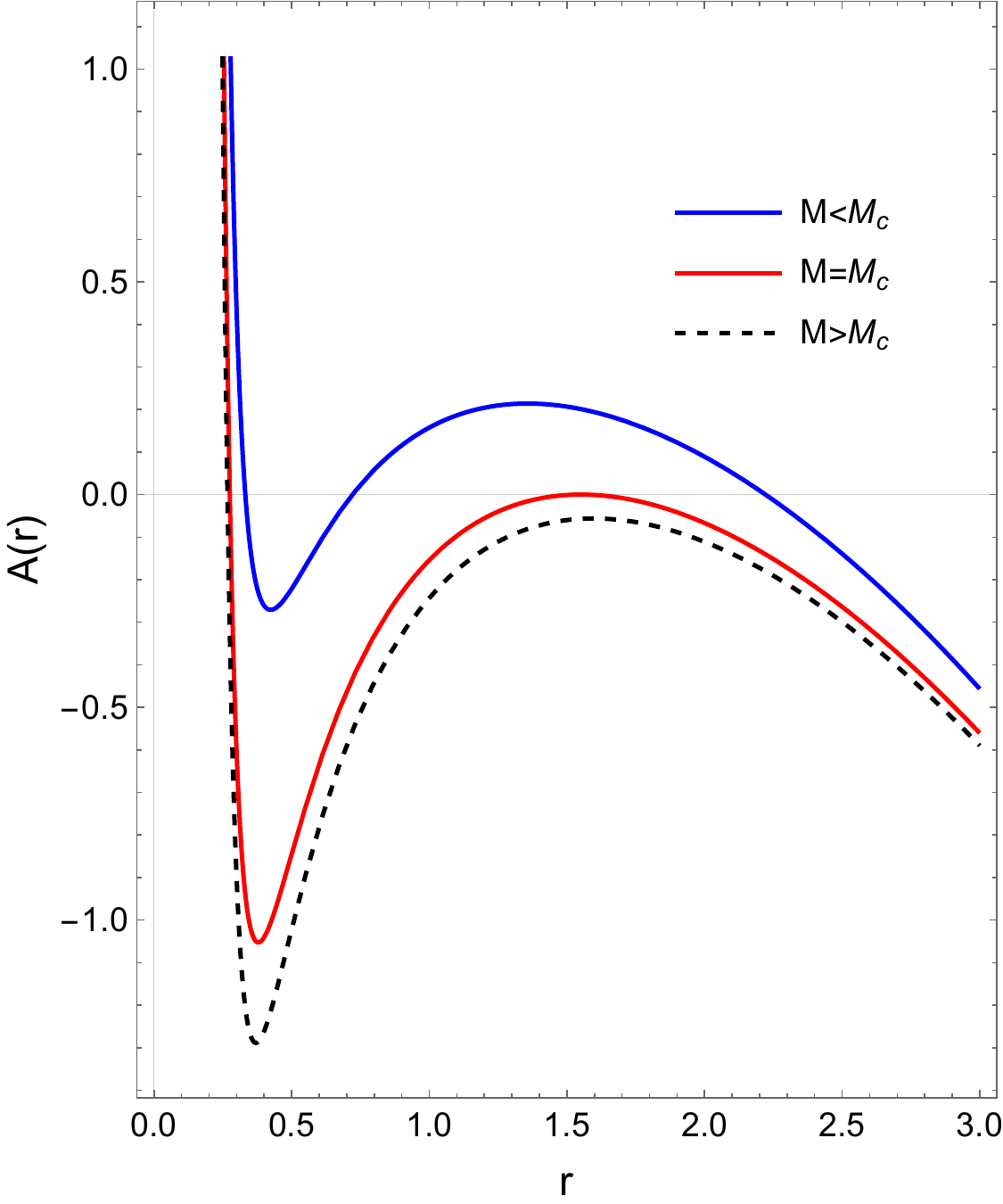}
	\hspace{1cm}
	\includegraphics[width=7cm, height=6cm]{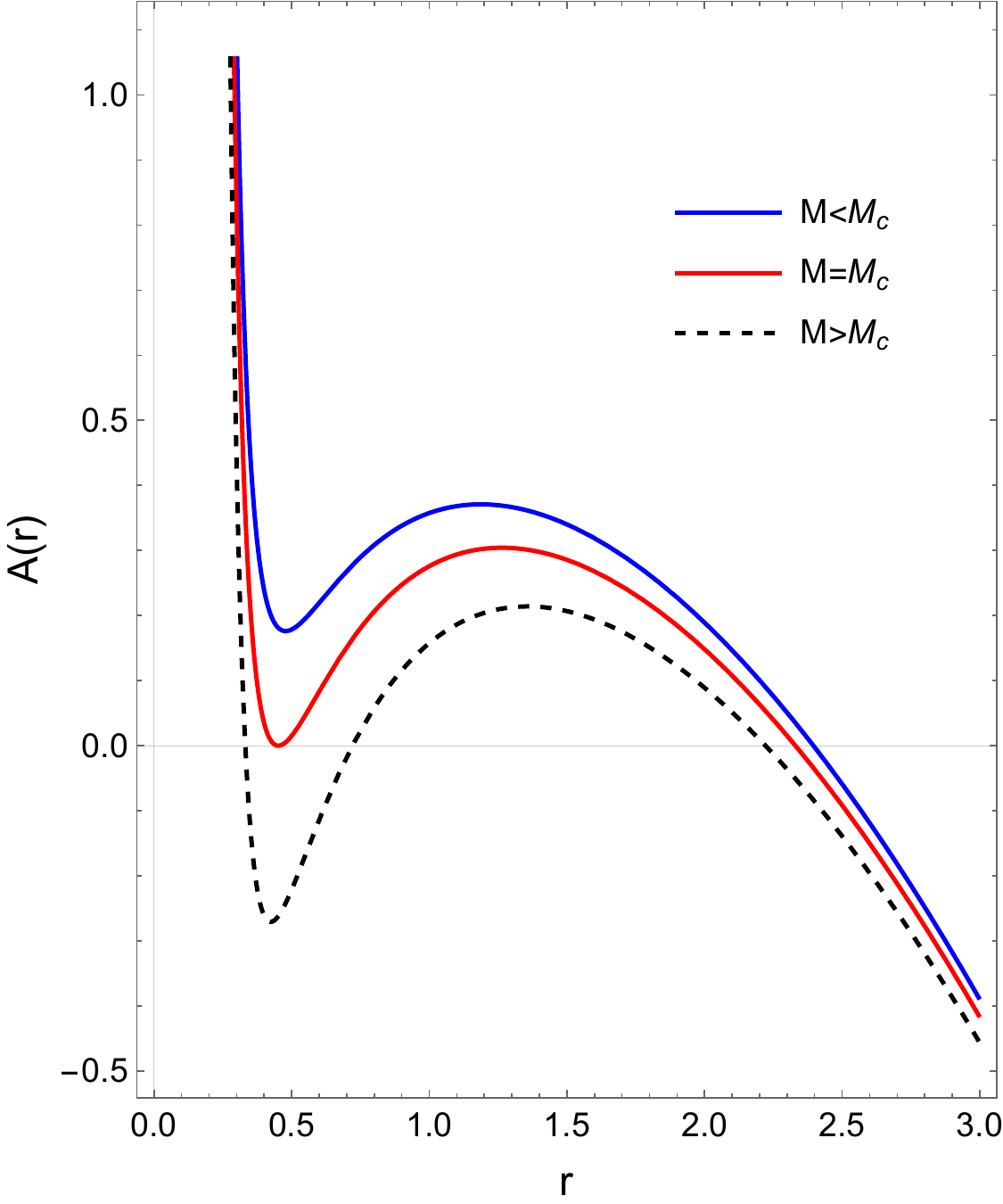}
	\caption{Graphical representation of the function $A(r)$ given in Eq. \eqref{eq:arexp} for the specific values: $q=0.3$, $\Lambda=0.4$, $f_k =2$, $a_0=1$. The left plot depicts the value $M_c  = 0.3408$; and the right plot $M_c = 0.5558$.}
	\label{fig:1}
\end{figure*}

To start, we propose a generalization of the Maxwell Lagrangian density, that is
\begin{equation}
    \mathcal{L}(F) = a_0 F^{f_k}+F,
    \label{eq:Lexp1}
\end{equation}
where $f_k$ is a dimensionless constant and $a_0$ has the unit $[L]^{4f_k-1}$.  Then we use Eq. \eqref{eq:Lexp1} in the Harada´s equations \eqref{eq:motion} and integrate to find 
\begin{equation}
    A(r) = 1-\frac{2 M}{r} + \frac{a_0 2^{1-f_k} q^{2 f_k} r^{2-4 f_k}}{4 f_k-3}+\frac{q^2}{r^2}-\frac{\Lambda  r^2}{3}.
    \label{eq:arexp}
\end{equation}
 This solution is a generalization of the Reissner-Nordström-Ads solution and returns to Schwarzschild if we use the limit of $f_k \rightarrow 1$, $a_0 \rightarrow 0$, $\Lambda \rightarrow 0$, and $q \rightarrow 0$.
Based on the metric function, the event horizon $r_H$ is determined by calculating
\begin{equation}
    A(r_H) = 0,
    \label{eq:con1}
\end{equation}
and this equation can give more than one result (solutions with multiple horizons). 
If there is more than one horizon, we look for an extremization condition by imosing the following
\begin{equation}
   \dfrac{d A(r_H)}{d r_H} = 0.
   \label{eq:con2}
\end{equation}

For simpler solutions, the two equations above can be solved analytically. However, for the solutions considered in this paper, we have solved this system numerically.
As a first example, we set the following values for the parameters: $q= 0.3, a_0 = 1$, $\Lambda=0.4$ and $f_k=2$. In this way, we simultaneously set the two conditions \eqref{eq:con1} and \eqref{eq:con2} for the case of an extreme solution by obtaining the values for $r_H$ and the critical mass $M_c$. In this particular scenario, we obtain $M_c = 0.3408$ and $M_c = 0.5558$, and the number of horizons depends directly on the values of these parameters. The plots in Fig. \ref{fig:1} show the behavior of the metric function \eqref{eq:arexp}, where the mass takes the values $M > M_c$, $M = M_c$, and $M < M_c$.
Note that the behavior of the function in relation to $M_c$ is reversed in the figures, whereas in the left plot of Fig. \ref{fig:1} we have 3 horizons for $M> M_c$, in the right plot we only have 1. These three horizons are the Cauchy horizon, the inner horizon, the event horizon, and the cosmological horizon.

\begin{figure*}[t!]
    \centering
   \includegraphics[width=7cm, height=6cm]{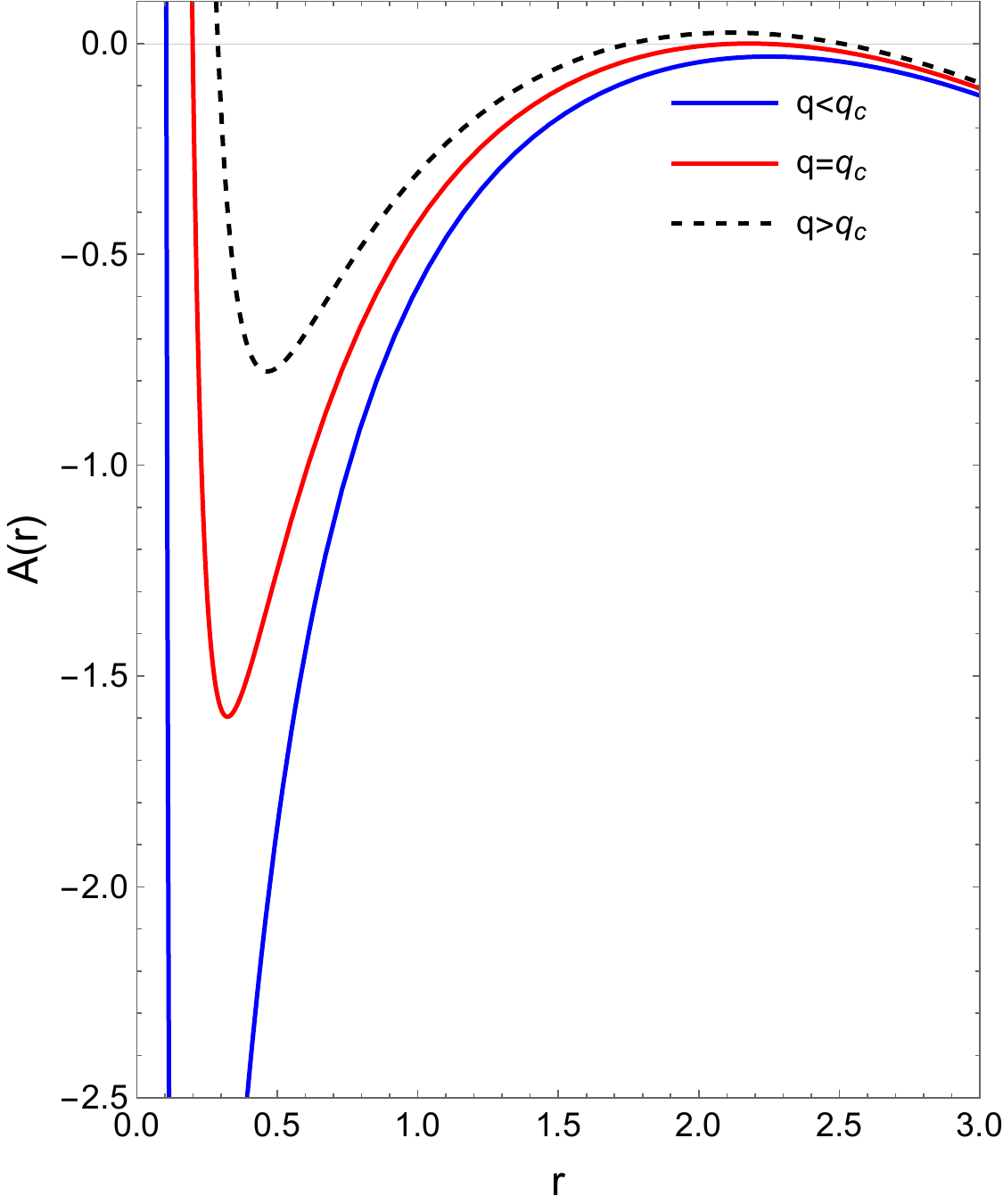}
   		\hspace{1cm}
   	\includegraphics[width=7cm, height=6cm]{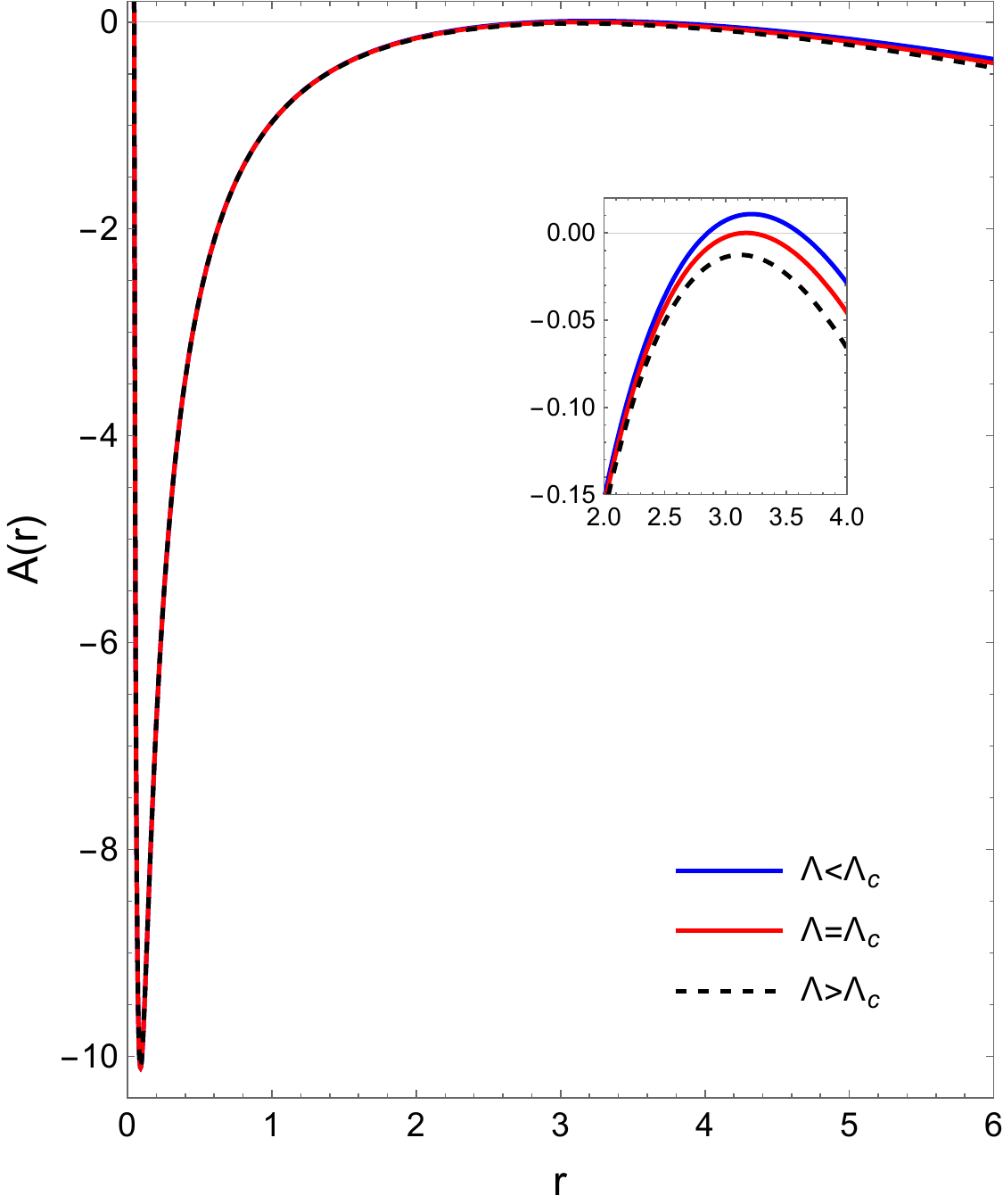}
    \caption{Graphical representation of the function $A(r)$ given in Eq. \eqref{eq:arexp}. The left plot depicts the specific values: $M=0.08$, $\Lambda=0.2$, $f_k =2$, $a_0=0.01$ and $q_c = 0.4912$. The right plot shows the following values: $q=0.3$, $M=1$, $f_k =0.1$, $a_0=0.1$ and $\Lambda_c = 0.006$.}
    \label{fig:3}
\end{figure*}

For the second numerical solution, we now use the following values for the constants: $M=0.08$, $\Lambda=0.2$, $f_k =2$ and $a_0=0.01$. This leads us to the critical charge value of
$q_c = 0.4912$. The behavior of $A(r)$ as a function of the radius is shown in the left plot of Fig. \ref{fig:3} for the cases $q > q_c, q = q_c$, and
$q < q_c$. If the charge is greater than the critical value, there are three horizons; if the charge is equal to $q_c$, there are two horizons, and for $q<q_c$
we only have one horizon. It is worth noting that this case differs from the first case in that we have only obtained one critical charge value.

\begin{figure*}[t!]
	\centering
	\includegraphics[width=7cm, height=6cm]{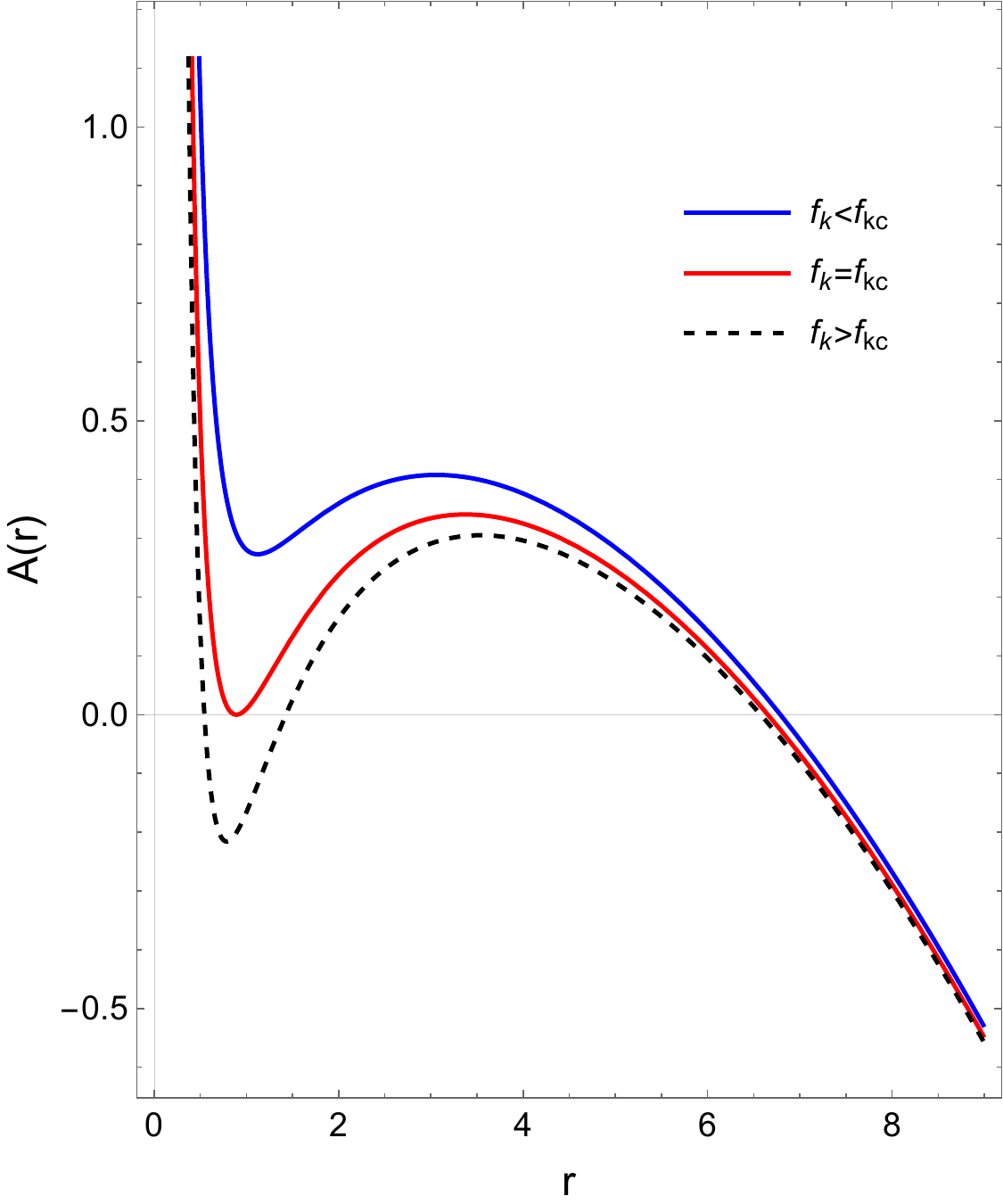}
	\hspace{1cm}
	\includegraphics[width=7cm, height=6cm]{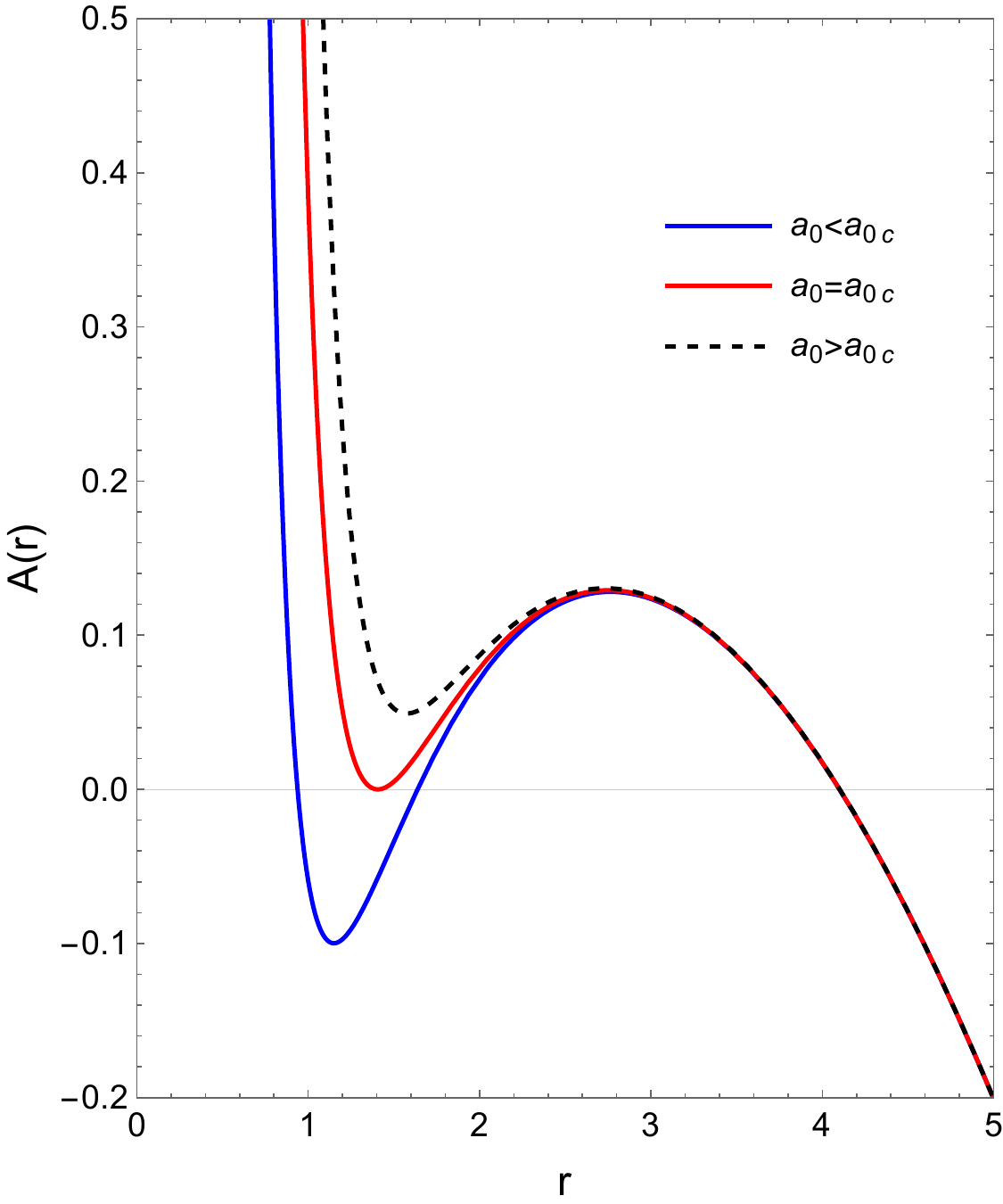}
	\caption{Graphical representation of the function $A(r)$ given in Eq. \eqref{eq:arexp}. In the left plot, we have the specific values: $q=0.6$, $\Lambda=0.05$, $M =1$, $a_0=1$ and $f_{kc} = 0.9081$. For the right plot, we have chosen the following specific values: $q=0.9$, $\Lambda=0.1$, $M =1$, $f_k=2$ and $a_{0c} = 9.2575$.}
	\label{fig:5}
\end{figure*}

If we now allow $\Lambda$ as a variable in Eqs. \eqref{eq:con1} and \eqref{eq:con2} and set the following values: $q=0.3$, $M=1$, $f_k =0.1$, $a_0=0.1$; we find a critical value $\Lambda_c = 0.006$. The right plot of Fig. \ref{fig:3} shows that  for $\Lambda < \Lambda_c$ we have three horizons, when $ \Lambda = \Lambda_c$ the second and the third horizon degenerate into just one, and for $\Lambda > \Lambda_c$ we only have one horizon.

To find a critical value for the exponent $f_k$, we set: $q=0.6$, $\Lambda=0.05$, $M =1$, $a_0=1$; and obtain
$f_{kc} = 0.9081$. In this case, as can be seen in the left plot of Fig. \ref{fig:5}, we have three horizons when $f_k > f_{kc}$, two for $f_k = f_{kc}$ and only one for $f_k < f_{kc}$. Finally, we choose: $q=0.9$, $\Lambda=0.1$, $M =1$, $f_k=2$; to look for a critical value of the parameter $a_0$. We find $a_{0c} = 9.2575$, and the behavior of the metric is depicted in the right plot of Fig. \ref{fig:5}. As we can see, for $a_0 >a_{ac}$, i.e. the solution has only one horizon.

\begin{figure}[t!]
    \centering
   \includegraphics[width=0.95\linewidth]{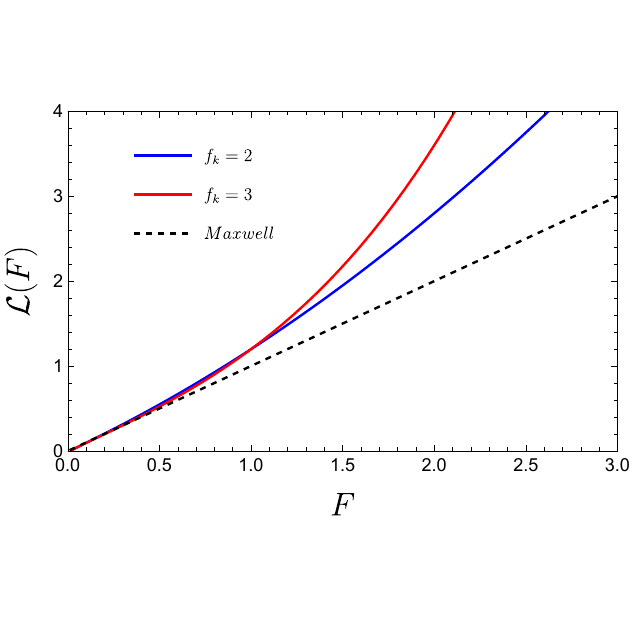}
    \caption{Graphical representation of the Lagrangian density \eqref{eq:Lexp} as a function of the Maxwell scalar $F$ for the values:  $q=0.3$, $\Lambda=0$, $f_0 =0$, $f_1 =0.1$  and $a_{0} = 0.2$.}
    \label{fig:Lexp}
\end{figure}

As a consistency check, we use the metric function \eqref{eq:arexp} in Eq. \eqref{eq:lagrangianageral} to obtain a Lagrangian density that also generates this solution. By doing so we get
\begin{equation}
    \mathcal{L}(F) = a_0 F^{f_k}-\frac{  q^3 f_1}{ \sqrt{2} \sqrt{F}}+f_0+F+\frac{\Lambda }{2}.
    \label{eq:Lexp}
\end{equation}
This new Lagrangian density brings with it a unique term that does not occur in the GR. The parameter $f_1$ is only associated with CKG, and therefore this Lagrangian density has a greater matter content and is of course also more general. 
This function is of Maxwell type in the limit of $F \rightarrow 0$, as we can see in Figure \ref{fig:Lexp}. We also note that increasing $f_k$ raises the value of the Lagrangian.
Furthermore, we can analyze the regularity of the spacetime generated by this geometry with the help of the Kretschmann scalar \cite{Lobo:2020ffi}. For this metric function, we have
\begin{eqnarray}
  K &=& 
    \frac{2 r^2 \zeta_2{}^2+4 (\zeta_1-1){}^2 \zeta_1{}^2}{r^4 \zeta_1{}^4}
    	+\frac{r^4 \zeta_3{}^2+2 r^2
   \zeta_2{}^2}{(\zeta_1-1){}^4}
   	\nonumber \\
   &&
   +\frac{\zeta_2{}^4}{\zeta_1{}^6} -\frac{2 \zeta_3 \zeta_2{}^2}{\zeta_1{}^5},
\end{eqnarray}
where
\begin{equation}
\begin{aligned}
    \zeta_1 = \frac{a_0 2^{1-f_k} q^{2 f_k} r^{2-4 f_k}}{4 f_k-3}-\frac{2 M}{r}+\frac{q^2}{r^2}-\frac{\Lambda  r^2}{3},
\end{aligned}
\end{equation}
\begin{equation}
    \zeta_2 = \frac{a_0 2^{1-f_k} (2-4 f_k) q^{2 f_k} r^{1-4 f_k}}{4 f_k-3}+\frac{2 M}{r^2}-\frac{2 q^2}{r^3}-\frac{2
   \Lambda  r}{3},
\end{equation}
\begin{eqnarray}
   \zeta_3 = \frac{a_0 2^{2-f_k} \left(8 f_k{}^2-6 f_k+1\right) q^{2 f_k} r^{-4 f_k}}{4 f_k-3}
   \nonumber \\ 
   -\frac{2 \Lambda}{3} -\frac{4 M}{r^3}+\frac{6 q^2}{r^4}.
\end{eqnarray}
This scalar is indeterminate in the limit $r \rightarrow 0$, indicating a curvature singularity at this point.

\subsection{Model 2}

For our second model, we consider
\begin{equation}
    \mathcal{L}(F) = F \left( \frac{1 +F^2}{1 +F} \right),
    \label{eq:lLinearGR}
\end{equation}
and consequently obtain the following solution
\begin{eqnarray}
	&&	A(r) =  1 -\frac{2 M}{r}  -\frac{\Lambda  r^2}{3} 	+\frac{\ 2^{3/4} q^{3/2} }{4 r} \left( 2 q^4-20 q^2 r^4 \right)
			\nonumber \\   
	&& 	+\frac{\ 2^{3/4} q^{3/2} }{4 r} \Bigg[\Biggl(\log \left(\frac{2\ 2^{3/4} \sqrt{q} r}{q-2^{3/4} \sqrt{q}
   r+\sqrt{2} r^2}+1\right)
			\nonumber \\
	&&	+2 \tan ^{-1}\left(1-\frac{2^{3/4} r}{\sqrt{q}}\right)-2 \tan
   ^{-1}\left(\frac{2^{3/4} r}{\sqrt{q}}+1\right)\Biggr) \Bigg]
	\label{eq:arl}
\end{eqnarray}

In this case, we also regain the Schwarzschild solution in the limits $\Lambda \rightarrow 0$ and $q \rightarrow 0$.
As before, we obtain a critical mass value $M_c$, for which we set the values $q=0.3$ and $\Lambda = 0.4$. Solving Eqs. \eqref{eq:con1} and \eqref{eq:con2} numerically, we obtain two values: $M_c=0.0876$ and $M_c = 0.3386$. This is because we have again used a positive value for the cosmological constant, so that the black hole has three horizons: The inner Cauchy horizon, the event horizon, and the exterior cosmological horizon.

\begin{figure*}[t!]
	\centering
	\includegraphics[width=7cm, height=6cm]{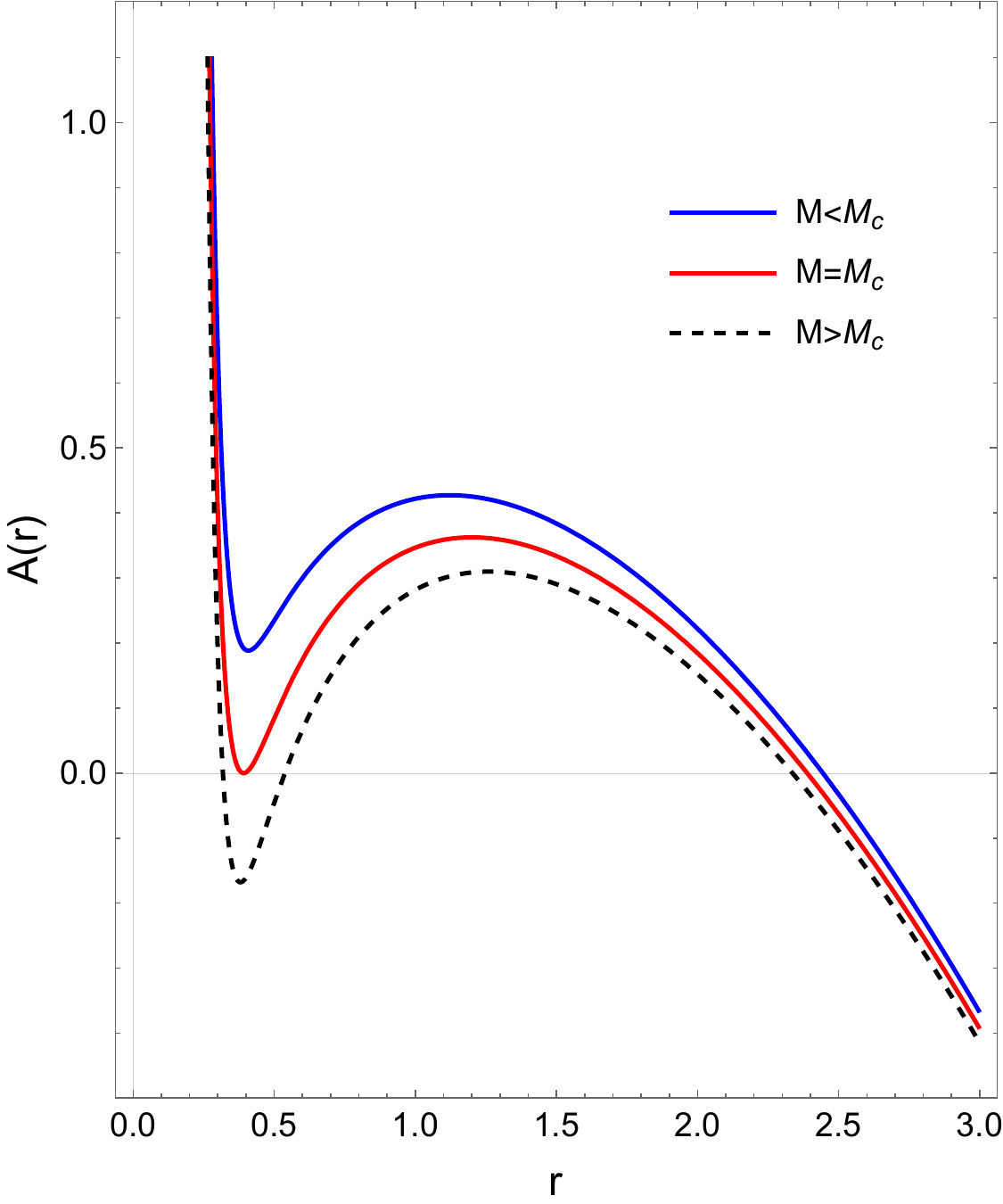}
	\hspace{1cm}
	\includegraphics[width=7cm, height=6cm]{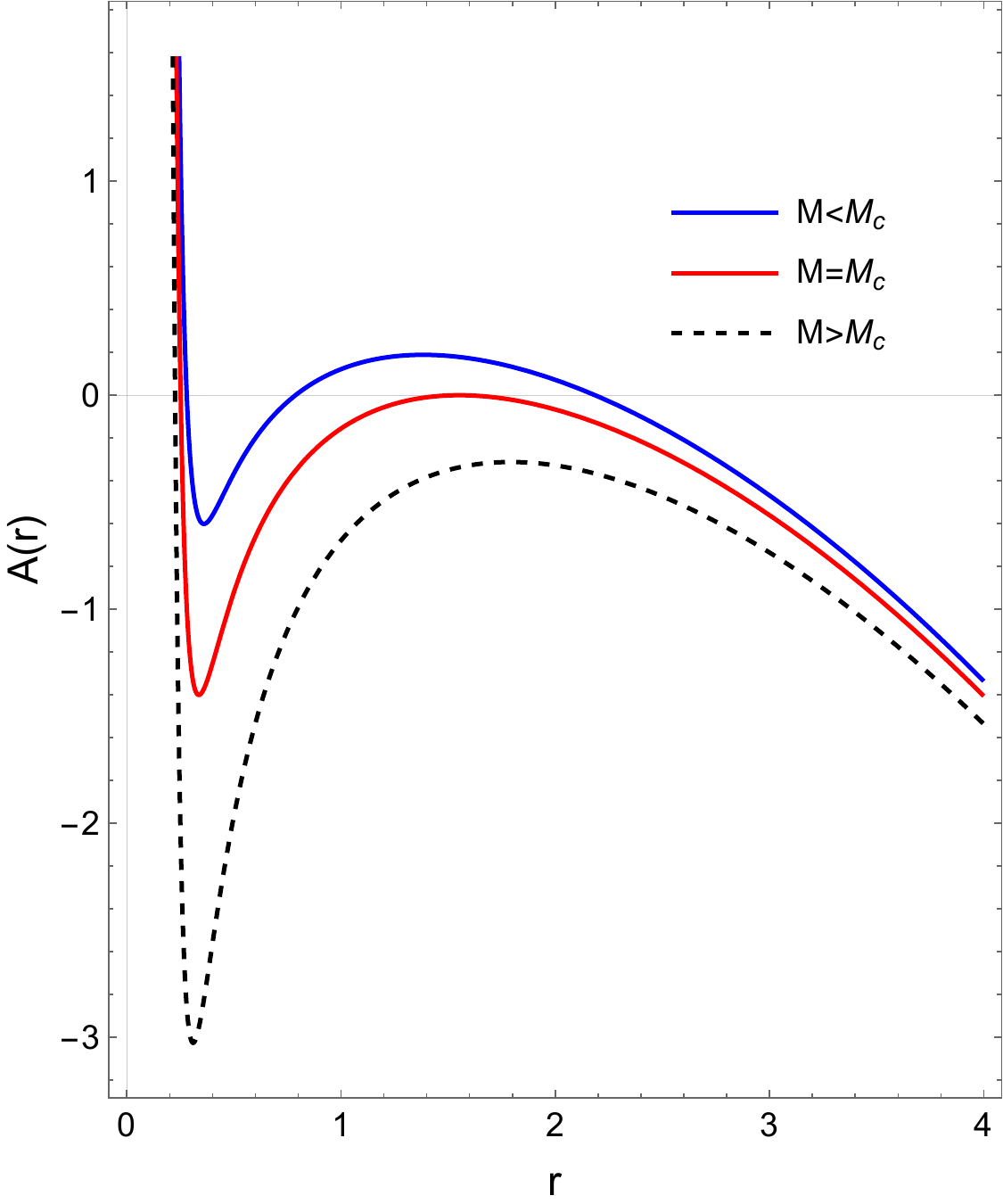}
	\caption{Graphical representation of the function $A(r)$ given in Eq. \eqref{eq:arl} for the specific values: $q=0.3$, $\Lambda=0.4$. The left plot depicts the value of $M_c=0.0876$; and the right plot $M_c=0.3386$.}
	\label{fig:7}
\end{figure*}

\begin{figure*}[t!]
	\centering
	\includegraphics[width=7cm, height=6cm]{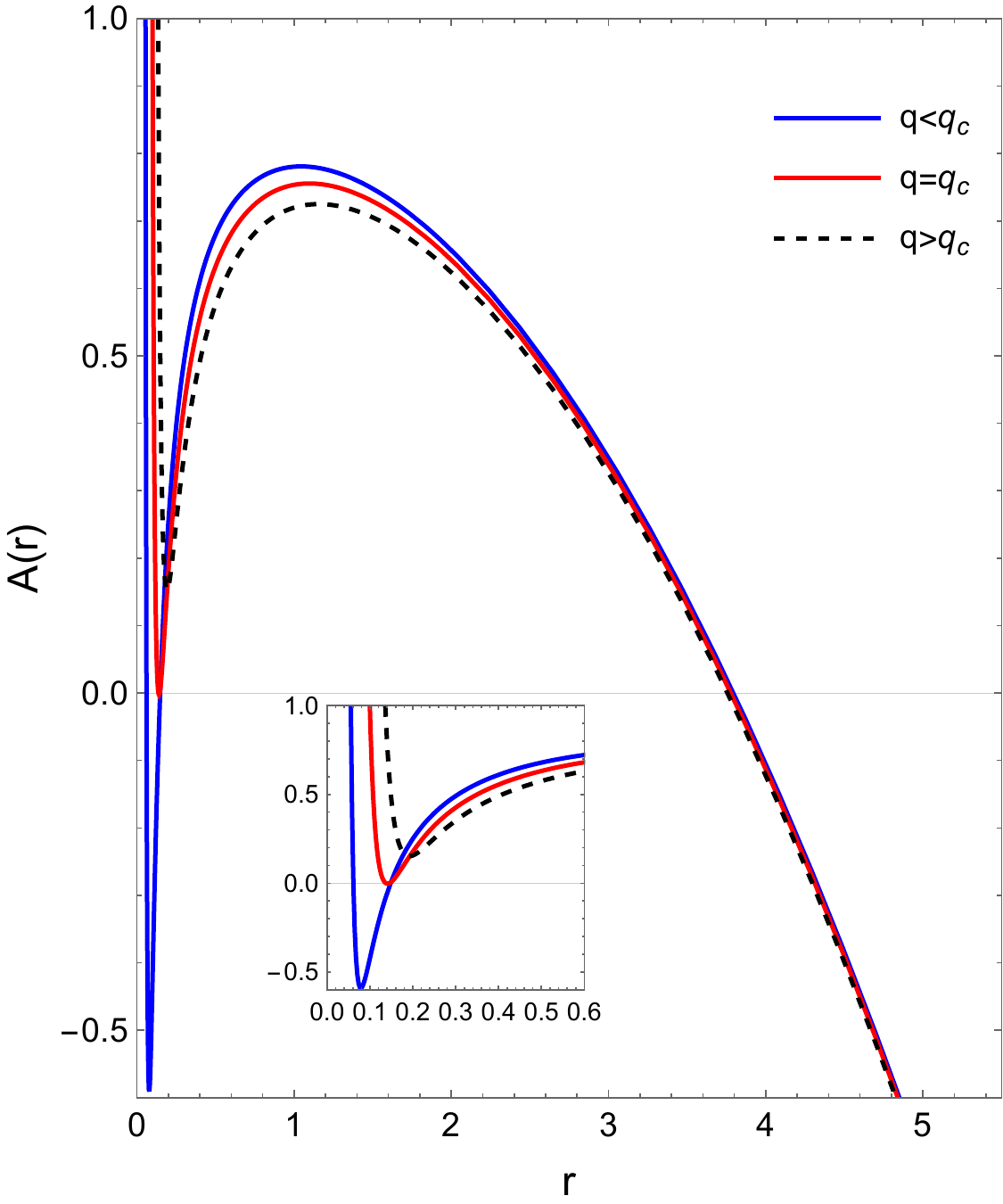}
	\hspace{1cm}
	\includegraphics[width=7cm, height=6cm]{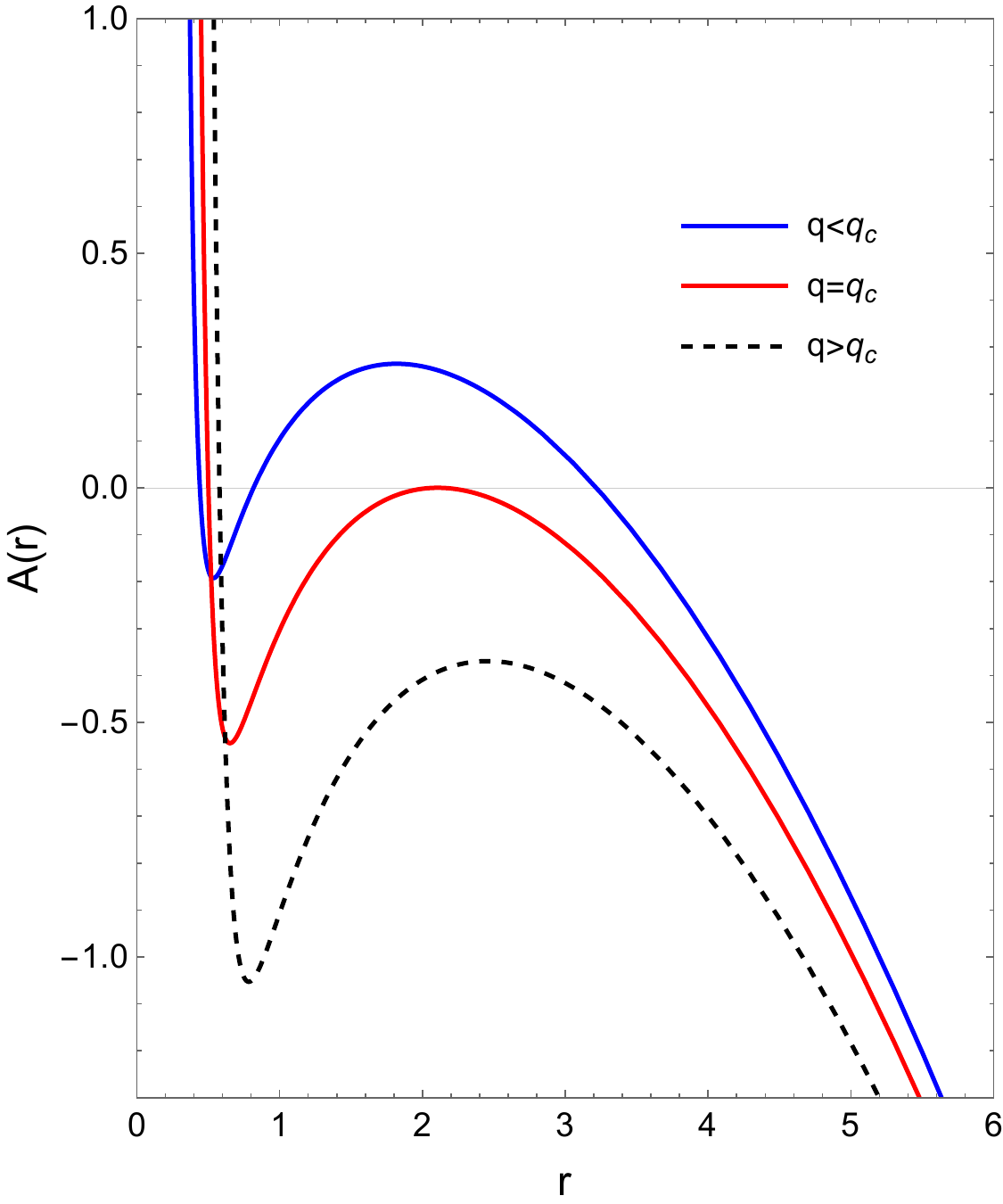}
	\caption{Graphical representation of the function $A(r)$ given in Eq. \eqref{eq:arl} for the specific values: $M=0.07$, $\Lambda=0.2$. The left plot depicts the value $q_c=0.0652$; and the rigth plot $q_c=0.7106$.}
	\label{fig:9}
\end{figure*}

As we can see in the left plot of Fig. \ref{fig:7}, the value $M_c = 0.0876$ limits the transition between the Cauchy horizon and the event horizon; in the right plot of Fig. \ref{fig:7}, the critical value $M_c = 0.3386$ represents the transition between the event horizon and the cosmological horizon. For both cases we have three horizons for $M>M_c$
two if the mass is critical $M= M_c$, and only one if $M< M_c$.

Next, we set the values: $M=0.07$ and $\Lambda=0.2$; and thus we find two critical charge values $q_c=0.0652$ and $q_c=0.7106$, as shown in the plots of Fig. \ref{fig:9}. We see that the black hole has three event horizons for $q<q_3$, two when $q=q_c$ and only one for $q>q_c$.

\begin{figure*}[t!]
    \centering
   \includegraphics[width=7cm, height=6cm]{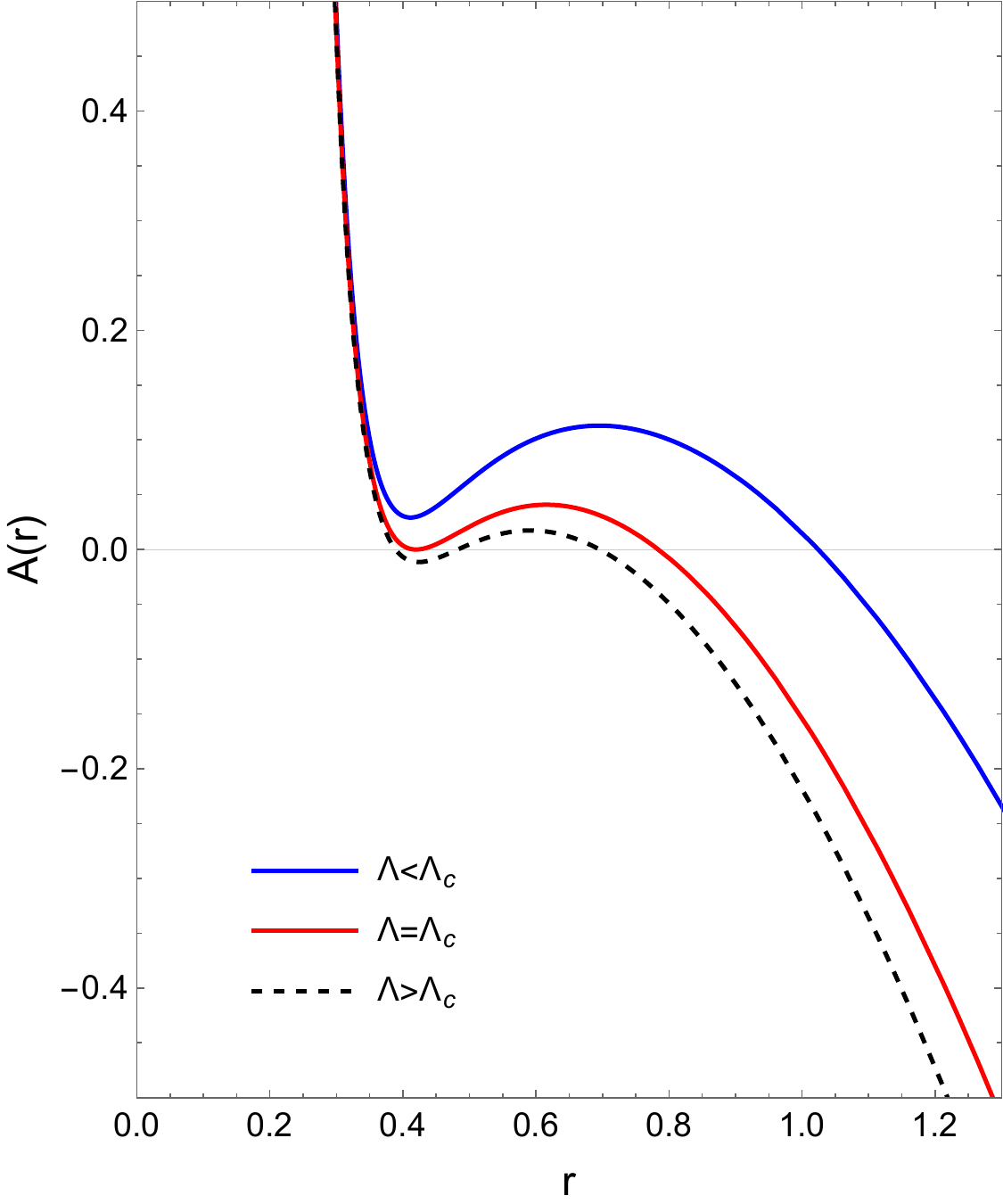}
   \hspace{1cm}
   \includegraphics[width=7cm, height=6cm]{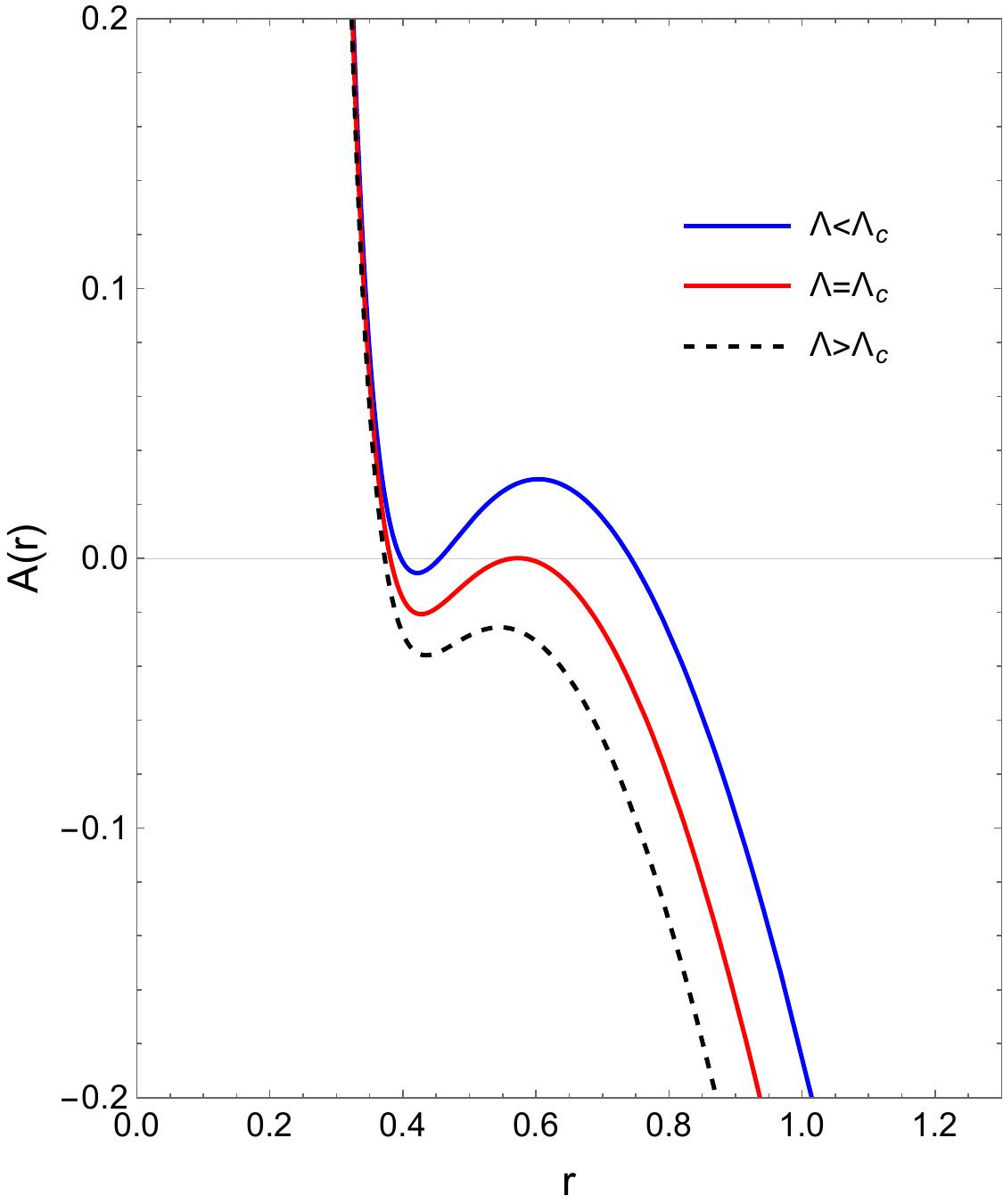}
    \caption{Graphical representation of the function $A(r)$ given in Eq. \eqref{eq:arl} for the specific values: $M=0.07$, $q=0.3$. The left plot depicts the value of $\Lambda_c=2.006$; and and the right plot $\Lambda_c=2.3539$.}
    \label{fig:11}
\end{figure*}

Finally, we choose: $M=0.07$ and $q=0.3$; and so we find $\Lambda_c=2.006$ and $\Lambda_c=2.3539$. In this case, when merging the first and second horizons, we have three horizons for $\Lambda > \Lambda_c$, as shown in the left plot of Fig. \ref{fig:11}. If, on the other hand, the second and third horizons are merged, the opposite is the case: there is only one horizon for $\Lambda > \Lambda_c$, as shown in the right plot of Fig. \ref{fig:11}.

\begin{figure}[t!]
    \centering
   \includegraphics[width=0.8\linewidth]{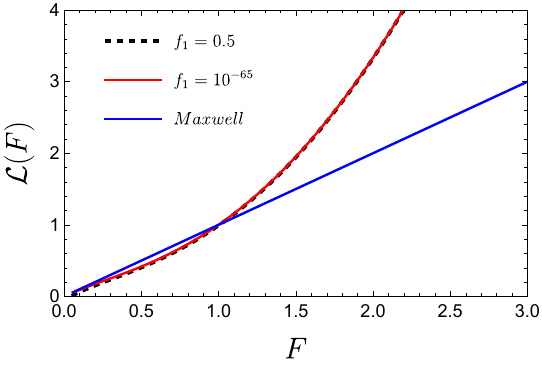}
    \caption{Graphical representation of the Lagrangian density \eqref{eq:Llinear} as a function of the Maxwell scalar $F$ for the specific values:  $q=0.3$, $f_0=0$ and $\Lambda=0$.}
    \label{fig:Llinear}
\end{figure}

To find the Lagrangian density that generates this solution in the CKG theory, we again insert the metric \eqref{eq:arl} into Eq. \eqref{eq:lagrangianageral}, which yields
\begin{equation}
    \mathcal{L}(F) =  F \left( \frac{1 +F^2}{1 +F}\right)- \frac{ 2q^3 f_1}{ \sqrt{2} \sqrt{F}}+\frac{\Lambda }{2}+f_0.
    \label{eq:Llinear}
\end{equation}
We verify again that the Lagrangian density in the CKG 
\eqref{eq:Llinear} is more general than that one 
\eqref{eq:lLinearGR} from GR.
If we now represent this function in Fig. \ref{fig:Llinear} by varying the values of the constant $f_1$, we see that it has little influence on the behavior of the Lagrangian. The reason why we use this particular value $f_1 = 10^{-65}$ will become clear in the next section.

To search for curvature singularities, we calculate the Kretschmann scalar.
\begin{widetext}
\begin{equation}
    \begin{aligned}
      & K = \frac{8 \Lambda ^2}{3}+\Biggl\{ 2 \bigl(8 q^8 r^{10} \left(75 M^2-580 M r^3+478 r^6\right) +239
   q^{16}  +320 q^4 r^{18}
   \left(45 M^2+22 M r^3+5 r^6\right)+19200 M^2 q^2 r^{22} \\ &
   +q^{12} \left(596 r^8-560 M
   r^5\right)-16 q^{10} r^9 \left(205 M+62 r^3\right) 
   +960 M q^6 r^{14} \left(5 M+3 r^3\right)+1152 q^{14} r^4\bigr)+9600 M^2 r^{26} \Biggr\}  \\ &
  \times \bigl[25 r^{16} \left(q^2+2 r^4\right)^4\bigr]^{-1} 
  +\Biggl\{ q^{3/2} \Biggl(\log
   \left(\frac{2\ 2^{3/4} \sqrt{q} r}{-2^{3/4} \sqrt{q} r+q+\sqrt{2} r^2}+1\right)
   +2 \tan ^{-1}\left(1-\frac{2^{3/4} r}{\sqrt{q}}\right)-2 \tan ^{-1}\left(\frac{2^{3/4}
   r}{\sqrt{q}}+1\right)\Biggr) \\& \times 
    \Biggl[8 \sqrt[4]{2} \biggl(13 q^6 r^4-60 M q^2 r^9-60
   M r^{13}+7 q^8 
   -q^4 \left(15 M r^5+22 r^8\right)\biggr) \\ &
   +15 q^{3/2} r^5 \left(q^2+2 r^4\right)^2 \Biggl(\log \left(\frac{2\
   2^{3/4} \sqrt{q} r}{-2^{3/4} \sqrt{q} r+q+\sqrt{2} r^2}+1\right) 
   +2 \tan ^{-1}\left(1-\frac{2^{3/4}
   r}{\sqrt{q}}\right) 
   -2 \tan ^{-1}\left(\frac{2^{3/4} r}{\sqrt{q}}+1\right)\Biggr)\Biggr]\Biggr\}  \\&
   \times \bigl[5 \sqrt{2}
   r^{11} \left(q^2+2 r^4\right)^2\bigr]^{-1} 
   -\frac{8 \Lambda  q^4 \left(q^4+4 q^2 r^4-4 r^8\right)}{3 r^8
   \left(q^2+2 r^4\right)^2},
    \end{aligned}
\end{equation}
\end{widetext}
and verify that this spacetime has a singularity at the origin $r=0$.

\subsection{Model 3}

The Lagrangian density of our last model is
\begin{equation}
    \mathcal{L} (F) = \sin (F),
\end{equation}
which, if integrated, results in
\begin{eqnarray}
         A(r) &=& 1- \frac{2M}{r} - \frac{\Lambda r^2}{3} - \frac{2r^2}{3} \sin \left(\frac{q^2}{2 r^4}\right)
        \nonumber  \\ 
         && - \frac{q^2}{6 r^2}
   \left[E_{\frac{3}{4}}\left(-\frac{i q^2}{2 r^4}\right)+E_{\frac{3}{4}}\left(\frac{i q^2}{2
   r^4}\right)\right].
    \label{eq:AsinF}
\end{eqnarray}

\begin{figure}[t!]
    \centering
   \includegraphics[width=7cm, height=6cm]{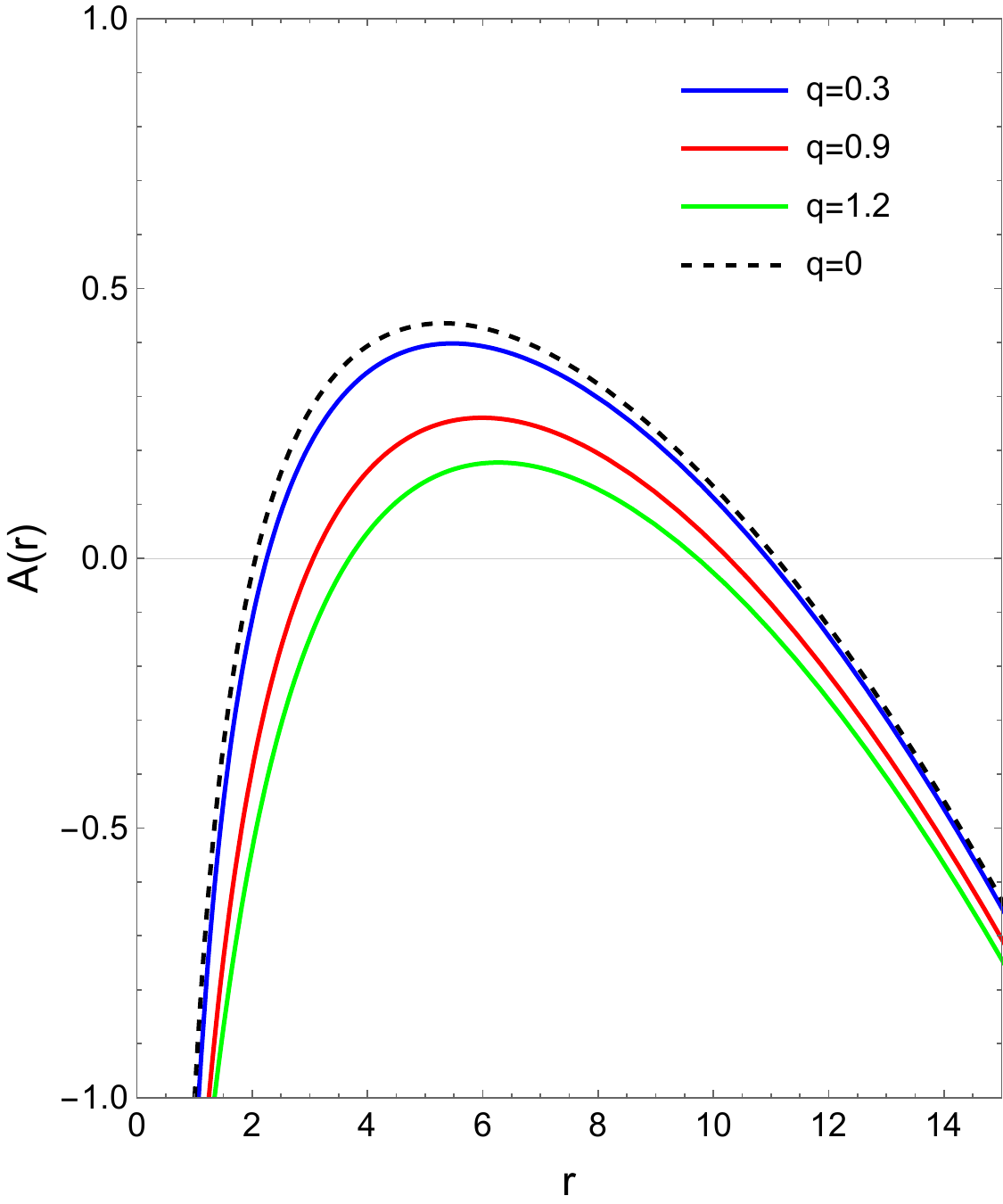}
    \caption{Graphical representation of the function $A(r)$ given in Eq. \eqref{eq:AsinF} for the specific values: $M=1$ and $\Lambda_c=0.02$.}
    \label{fig:13}
\end{figure}

\begin{figure*}[t!]
    \centering
   \includegraphics[width=7cm,height=6cm]{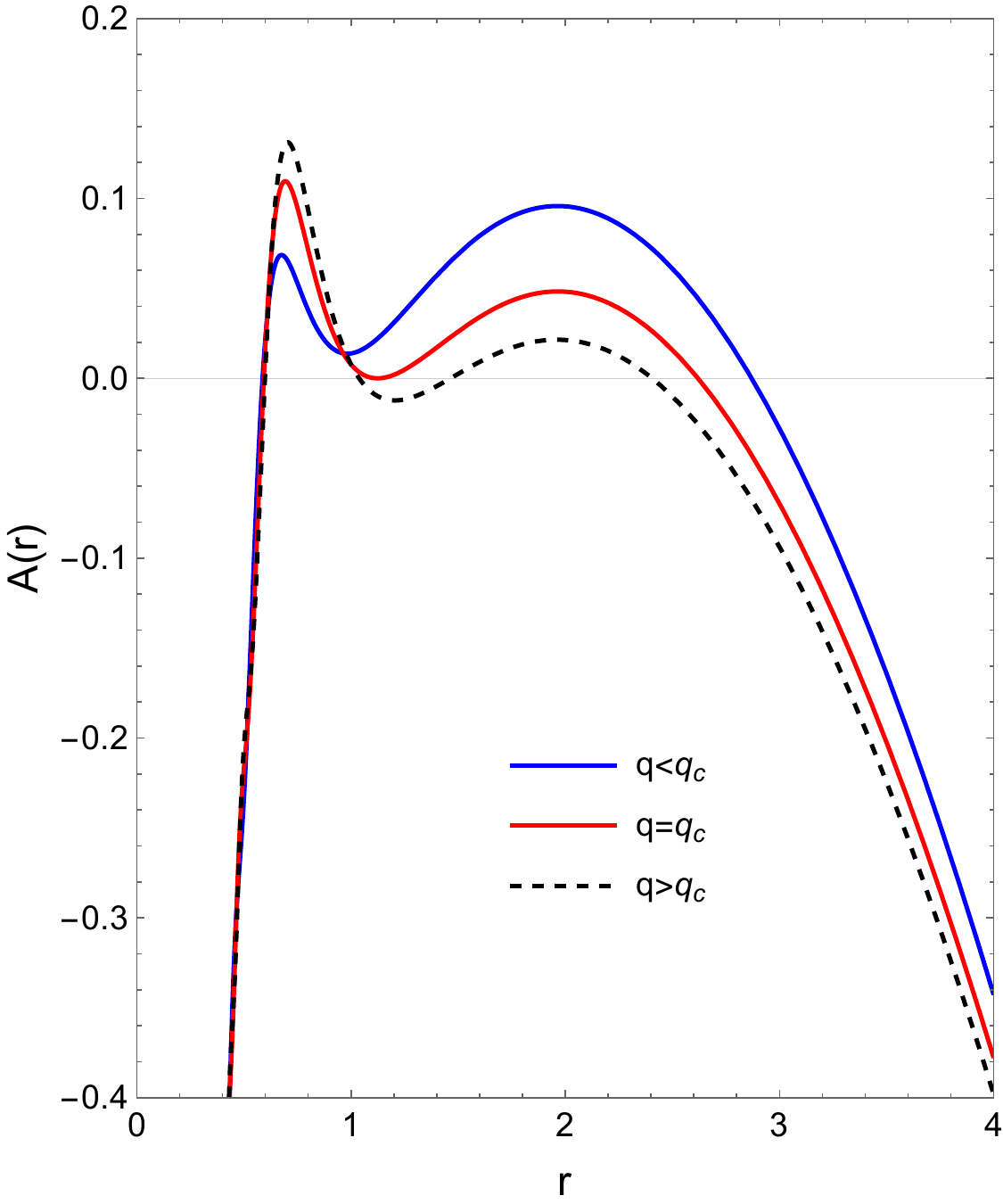}
    \hspace{1cm}
   \includegraphics[width=7cm,height=6cm]{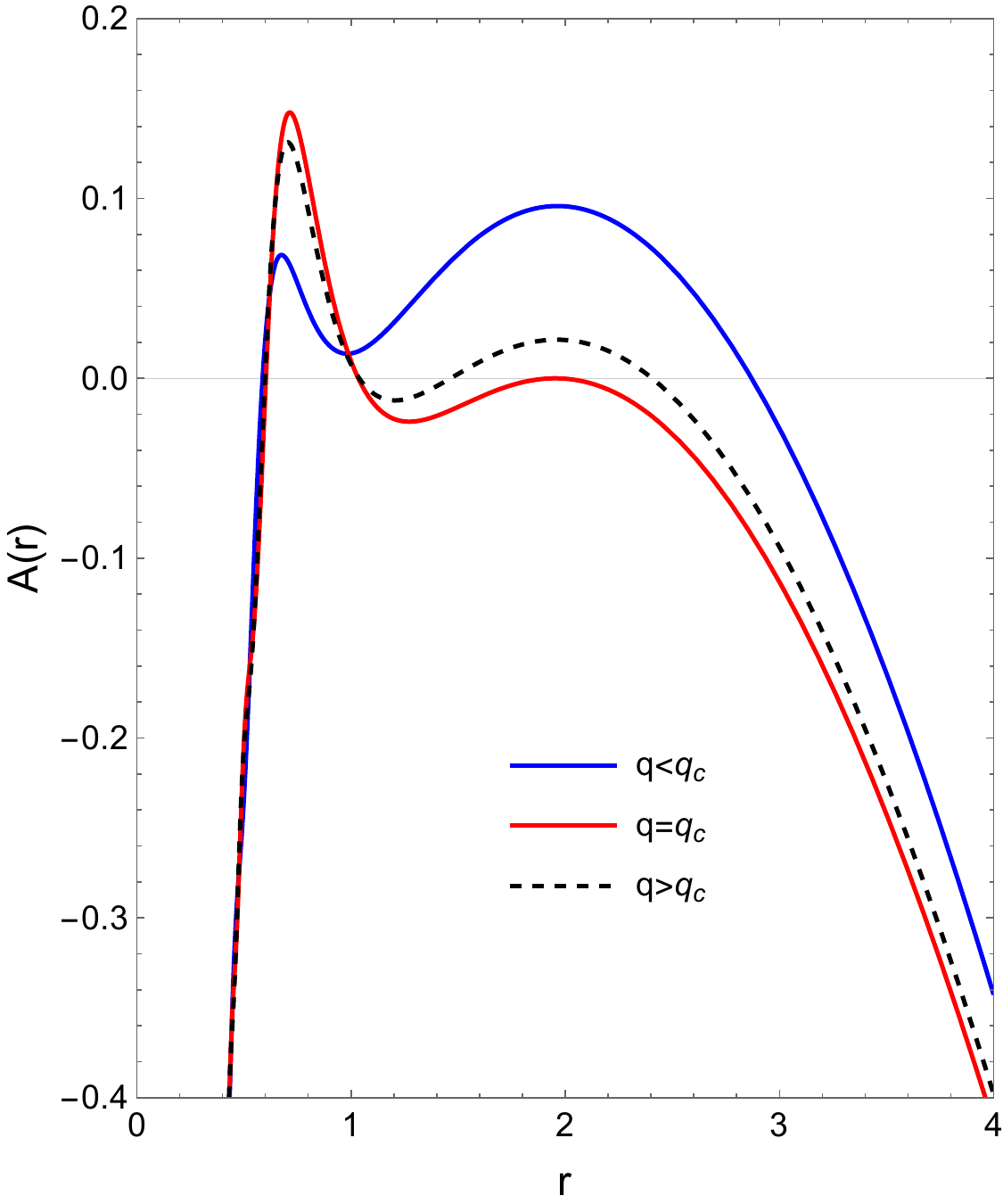}
    \caption{Graphical representation of the function $A(r)$ given in \eqref{eq:AsinF} for the specific values: $M=1$ and $\Lambda_c=0.02$. The left plot depicts the value $q_c = 0.9958$; and the right plot the value $q_c = 1.0934$.}
    \label{fig:14}
\end{figure*}

Note that for $q= \Lambda = 0$ the solution is reduced to the standard Schwarzschild metric.
The sinusoidal behavior of the function \eqref{eq:AsinF} implies that the solution may possess multiple horizons. However, for values of $q \approx M$, the solution has only two horizons, as shown in Fig. \ref{fig:13}.
If we reduce the mass, we can construct a solution with three horizons and look for extremization conditions, as we did in the previous solutions. For example, if we set the values $M = 0.3$ and $\Lambda= 0.18$, we get two critical load values $q_c = 0.9958$ and $q_c = 1.0934$. The plots in Fig. \ref{fig:14} show this scenario.

\begin{figure}
    \centering
   \includegraphics[width=0.8\linewidth]{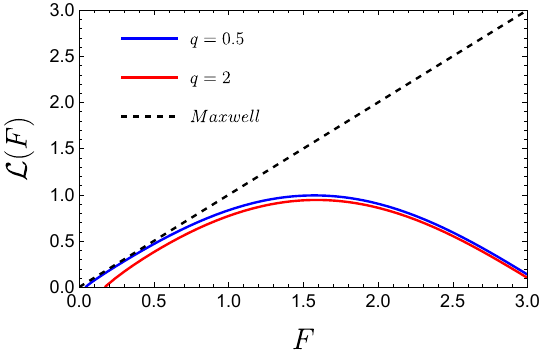}
    \caption{Graphical representation of the Lagrangian density given in \eqref{eq:Lsin} as a function of the Maxwell scalar $F$ for the specific values: $f_1=\Lambda=0$.}
    \label{fig:Lsin}
\end{figure}

The Lagrangian density that corresponding to CKG theory for this model is
\begin{equation}
   \mathcal{L}(F) = f_0+\frac{q \left(\lambda -2 q^2 f_1\right)}{2 \sqrt{2} \sqrt{F}}+\sin (F)+\frac{\Lambda }{2},
   \label{eq:Lsin}
\end{equation}
whose behavior is shown in Figure \ref{fig:Lsin}. We now vary the charge and find that a larger value of $q$ reduces the Lagrangian density.

Finally, we calculate the Kretschmann scalar, which is given by
\begin{widetext}
\begin{equation}
    \begin{aligned}
       & K = \Bigg\{8 \Biggl(2 q^6 \left(E_{-\frac{5}{4}}\left(-\frac{i q^2}{2 r^4}\right)+E_{-\frac{5}{4}}\left(\frac{i
   q^2}{2 r^4}\right)\right) \left(-30 M r^9+q^6 \left(E_{-\frac{5}{4}}\left(-\frac{i q^2}{2
   r^4}\right)+E_{-\frac{5}{4}}\left(\frac{i q^2}{2 r^4}\right)\right)\right) +4 r^4 \sin \left(\frac{q^2}{2
   r^4}\right)  \\ &
   \times \left(4 q^{10} \left(E_{-\frac{5}{4}}\left(-\frac{i q^2}{2
   r^4}\right)+E_{-\frac{5}{4}}\left(\frac{i q^2}{2 r^4}\right)\right)-60 M q^4 r^9+25 \Lambda 
   r^{20}\right)+60 q^8 r^8 \left(E_{-\frac{5}{4}}\left(-\frac{i q^2}{2
   r^4}\right)+E_{-\frac{5}{4}}\left(\frac{i q^2}{2 r^4}\right)\right)  \\&
 \times  \cos \left(\frac{q^2}{2
   r^4}\right)+r^8 \biggl(25 r^{10} \left(18 M^2+\left(\Lambda ^2+2\right) r^6\right)-50 q^2 r^9 \left(18
   M+\Lambda  r^3\right) \cos \left(\frac{q^2}{2 r^4}\right)+16 q^8+275 q^4 r^8 \\ &
   +\left(-16 q^8+275 q^4
   r^8-50 r^{16}\right) \cos \left(\frac{q^2}{r^4}\right)\biggr)+10 q^2 r^{12} \left(12 q^4-5 r^8\right)
   \sin \left(\frac{q^2}{r^4}\right)\Biggr)\Bigg\} \times (5 r^{24})^{-1},
    \end{aligned}
\end{equation}
\end{widetext}
and note that this model also has a curvature singularity at $r = 0$.

\section{Shadow}

\subsection{Shadow's radius \label{sec:rs}}

The shadow of a black hole is a silhouette, a region characterized by a sharp decrease in observed luminosity, projected onto the plane of the sky of a distant observer. This projection is created by photons that originate from a background source and are absorbed due to its strong gravity. To calculate the shadow of the black hole, we use the well-established formalism described in \cite{Perlick:2021aok}. In this approach, they consider a static observer located at finite distance ($r_0$, $\theta_0=\pi/2$) and projecting an angle $\theta$ relative to the radial coordinate $r$, as given by
\begin{equation}
    \tan\theta=\sqrt{A(r)C(r)}\frac{d\phi}{dr}\Bigg|_{r=r_{0}}.\label{tan}
\end{equation}
The shadow of the black hole, which is understood as the region bordering the critical curve (on the observer's plane image), is thus formed by the light rays radiated back to the source by the observer. The radius of the shadow is
\begin{equation}
r_{s}=r_{p}\sqrt{\frac{A(r_{0})}{A(r_{p})}},\label{r_shadow2}
\end{equation} 
where $r_p$ is the radius of the photon sphere  \cite{Claudel:2000yi}
\begin{equation}
    \frac{C^{\prime}(r_p)}{C(r_p)} = \frac{A^{\prime}(r_p)}{A(r_p)}\,.
    \label{eq:fotonesfera}
\end{equation}
The black hole distance and mass estimates reported in Table \ref{tab:my-table}.
\begin{table}[ht!]
\centering
\resizebox{\linewidth}{!}{%
\def\arraystretch{1.4}
\begin{tabular}{cccc}
\hline \hline
\multicolumn{4}{c}{Parameter values}                                                                               \\ \hline
\hspace{1 mm}Survey\hspace{1 mm} & \hspace{1 mm}$M (\times 10^6 M_{\odot})$\hspace{1 mm} & \hspace{1 mm}$D$\hspace{1 mm} (kpc) & \hspace{1 mm}Reference\hspace{1 mm} \\ 
Keck                             & $3.951 \pm 0.047$                                     & $7.953 \pm 0.050 \pm 0.032$         & {\cite{Do:2019txf}} \\ \hline
\end{tabular}%
}
\caption{Sgr A* Mass and distance.}
\label{tab:my-table}
\end{table}


However, since we have obtained solutions to the Einstein equations coupled with nonlinear electrodynamics, the light rays now follow a geodesic described by the effective metric~\cite{Novello:1999pg,Toshmatov:2021fgm}. This leads to a change in the photon sphere, which of course also results in a significant change in the shadow radius \eqref{r_shadow2}.
It is also worth noting that this approach is not always used in the literature. For instance, in \cite{Vagnozzi:2022moj}, the authors used the usual metric to treat the solution of Bardeen and others that can also be described by NED. The rationale for this is that NED is not the only way to interpret these solutions. However, since we are interested here precisely in these non-linear corrections, we will use the formalism of the effective metric, which is given by
\begin{equation}
     g_{eff}^{\mu\nu}={\cal L}_{F}g^{\mu\nu}-{\cal L}_{FF}F_{\sigma}^{\phantom{\sigma}\mu}F^{\sigma\nu},\label{g_efet}
 \end{equation}
 where ${\cal L}_{FF} = \partial {\cal L}_{F}/ \partial F$.
Since we consider a magnetic charge,  the new metric functions $\bar{A}(r)$ and $\bar{C}(r)$ are
\begin{equation}
\begin{aligned}
 & \bar{A}(r) = \frac{A(r)}{\cal L_F }, \\
  &\bar{C}(r) = \frac{C(r)}{{\cal L}_{F}+2F{\cal L}_{FF}}.
  \end{aligned}
   \label{eq:abar}
\end{equation}

We now try to narrow down the parameters of the solutions by checking the compatibility of the corresponding shadow radius $r_s$ of their geometry with the shadow size of Sgr A* derived from the EHT using the above data. This is because, as pointed out in \cite{EventHorizonTelescope:2022xqj}, the observed shadow size of Sgr $A^{\star}$ can be derived from the size of the bright emission ring, subject to a calibration factor that multiplicatively accounts for the theoretical and observed uncertainties. Such a correlation is possible thanks to the priorities for the mass-distance ratio of Sgr $A^{\star}$ obtained from measurements of the orbital dynamics of the stars closest to the Galactic center (i.e. the so-called S stars), listed in Table \ref{tab:my-table}. These measurements allow the EHT collaboration to determine the fraction of the deviation $\delta$ between {} inferred shadow size $r_{s}$ of Sgr $A^{\star}$ and that of a Schwarzschild black hole $r_{s,Sch}= 3 \sqrt{3} M$, with the angular radius $\omega_g = M/D$ via the formula
\begin{equation}\label{eq:fractionaldeviation}
    r_{s}/M=(\delta +1) 3 \sqrt{3} \ ,
\end{equation}
which, assuming a normal distribution of the estimation uncertainties, leads to the following shadow size limits \cite{Vagnozzi:2022moj}
\begin{equation} \label{eq:1sigma}
    4.55 \lesssim r_{s}/M \lesssim 5.22 \ ,
\end{equation}
at $1\sigma$ deviation and
\begin{equation} \label{eq:2sigma}
    4.21 \lesssim r_{s}/M \lesssim 5.56 \ ,
\end{equation}
at $2\sigma$ deviation.

\subsection{Model 1}

For our first model, we substitute Eq. \eqref{eq:arexp} in \eqref{eq:abar}, with $C(r) = r^2$, and have
\begin{equation}
 \bar{A}(r) = \frac{1-\frac{2 M}{r}+\frac{q^2}{r^2}-\frac{\Lambda
    r^2}{3}+\frac{2^{1-\text{fk}} a_0 q^{2 f_k} r^{2-4 f_k}}{4 f_k-3}}{a_0 2^{1-f_k} f_k \left(\frac{q^2}{r^4}\right)^{f_k-1}+r^6 f_1+1},
    \label{eq:arexpeff}
\end{equation}
\begin{equation}
	\bar{C}(r) =\left(\frac{r^2}{q^2} f_0 2^{1-f_k} f_k (2 f_k-1) \left(\frac{q^2}{r^4}\right)^{f_k}-2 r^4
		f_1+\frac{1}{r^2}\right)^{-1}.
\end{equation}

\begin{figure*}[t!]
	\centering
	\includegraphics[scale=0.65]{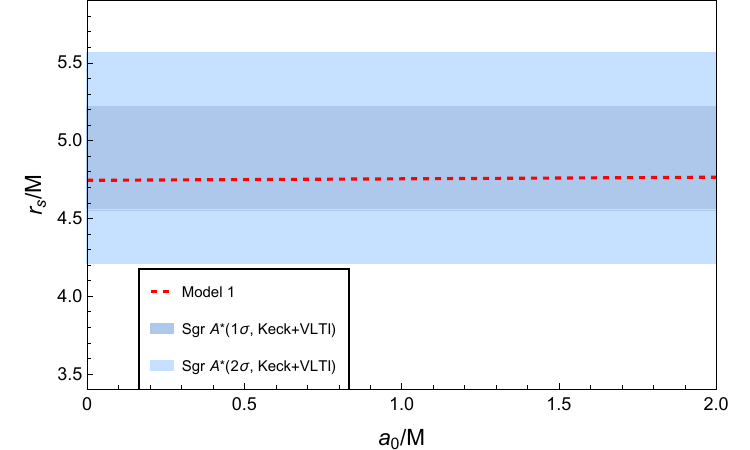}
	\hspace{1cm}
	\includegraphics[scale=0.65]{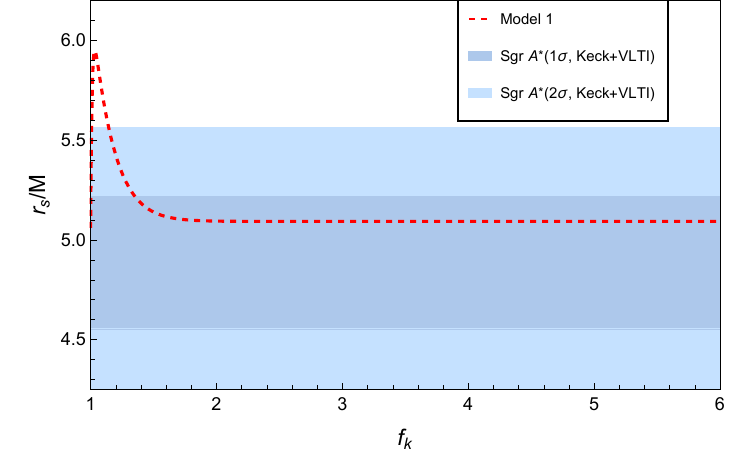}
	\caption{Shadow radius $r_s$ of the model of a black hole (red curve) metric function given \eqref{eq:arexpeff} and in units of BH mass $M$, as a function of $a/M$. In the left plot we have used the values: $f_k=2$, $q=0.5$, $\Lambda= 10^{-41}$ and $f_1= 2 \times 10^{-65}$. The right plot depicts the fixed values: $q=0.3$, $a_0 =0.5$, $f_1 = 2 \times 10^{-66}$ and $\Lambda = 10^{-52}$.
		The dark blue and light blue regions match the EHT horizon image of Sgr $A^{\star}$ to
		$1\sigma$ and $2\sigma$ respectively after averaging the Keck and VLTI mass-distance ratio priors for Sgr $A^{\star}$. The white regions are instead excluded by the same observations at more than $2 \sigma$.}
	\label{fig:16}
\end{figure*}

With this correction of NED, we see that the effective metric also depends on the constant $f_1$. So we could first try to restrict it to 5 parameters, but in \cite{Vagnozzi:2022moj} the authors have shown, using the Kottler solution, that the shadow radius with the EHT measurements for values of the cosmological constant within the interval $10^{-55} < \Lambda <10^{-40}$. Since the dependence of this model 1 and the other models is the same, i.e. $\Lambda r^2$, we will use this interval as a reference. Moreover, we obtained our results by numerically calculating the shadow radius, since it was impossible to solve the Eq. \eqref{eq:fotonesfera} for general values of the constants. We used the distance from the observer $r_0$ given in the Table \ref{tab:my-table} and fixed the mass $M=1$ for simplicity (so that the other quantities are given with respect to the mass). First we choose the values: $f_k=2$, $q=0.5$, $\Lambda= 10^{-41}$ and $f_1= 2 \times 10^{-65}$. The left plot of Fig. \ref{fig:16} depicts the shadow radius of model 1 as a function of $a_0/M$, we can see that $r_s$ increases with the parameter $a_0$. Next, we use: $q=0.3$, $a_0 =0.5$, $f_1 = 2 \times 10^{-66}$ and $\Lambda = 10^{-52}$; and the result is shown in the right plot of Fig. \ref{fig:16}, where we find that for $2>f_k>6$ the shadow radius is within the limit $1 \sigma$. An important remark is that $r_s$ is not well defined for $f_k<1$ and therefore we do not use these values in the graph. 

\begin{figure*}[t!]
    \centering
   \includegraphics[scale=0.65]{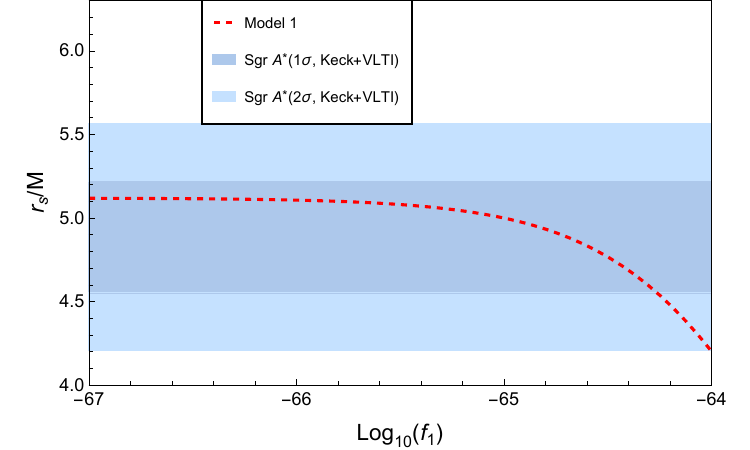}
   		\hspace{1cm}
    \includegraphics[scale=0.65]{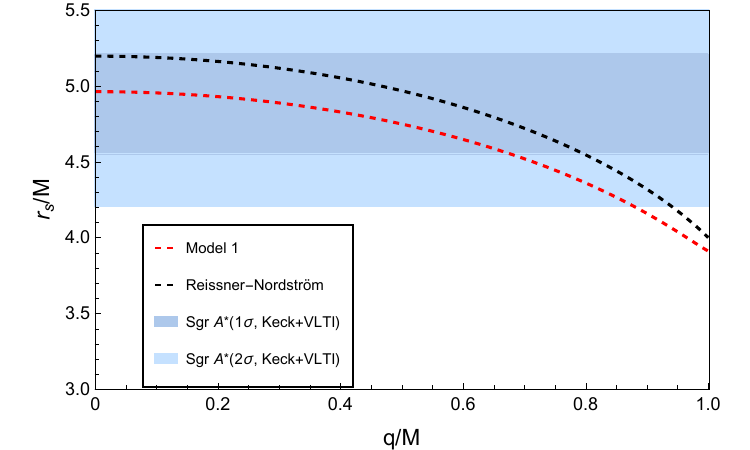}
    \caption{Same as Figure \ref{fig:16} for the effective metric of Eq. \eqref{eq:arexpeff}. The left plot depicts the fixed values: $q=0.3$, $a_0 =0.5$, $f_k = 2 $ and $\Lambda = 10^{-52}$. In the right plot, we consider: $f_k=2$, $a_0 =0.5$, $f_1 = 2 \times 10^{-66}$ and $\Lambda = 10^{-41}$.}
    \label{fig:17}
\end{figure*}

In the left plot of Fig. \ref{fig:17}, we vary the parameter $f_1$ in the interval $10^{-67} <f_1 < 10^{-64}$ and notice that increasing this constant decreases the radius of the shadow. 
Finally, we fix: $f_k=2$, $a_0 =0.5$, $f_1 = 2 \times 10^{-66}$ and $\Lambda = 10^{-41}$; and we calculate $r_s$ as a function of the normalized magnetic charge $q/M$, as shown in the right plot of Fig. \ref{fig:17}. The generated shadow is similar to the RN solution and also decreases with the addition of $q$. From this analysis, we conclude that the EHT observations are consistent with Sgr $A^{\star}$ being a black hole described by Model 1.

\subsection{Model 2}

Now, using Eqs. \eqref{eq:arl} and \eqref{eq:abar}, with $C(r) = r^2$, we have for the second model
\begin{widetext}
\begin{equation}
    \begin{aligned}
        & \bar{A}(r) =  \Biggl\{\left(q^2+2 r^4\right)^2 \Biggl[-20 r^5 \left(6 M+\Lambda  r^3-3 r\right)+15\ 2^{3/4} q^{3/2} r^5
   \Biggl[\log \left(\frac{2\ 2^{3/4} \sqrt{q} r}{-2^{3/4} \sqrt{q} r+q+\sqrt{2} r^2}+1\right)+2 \tan
   ^{-1}\left(1-\frac{2^{3/4} r}{\sqrt{q}}\right) \\ 
  & -2 \tan ^{-1}\left(\frac{2^{3/4}
   r}{\sqrt{q}}+1\right)\Biggr]+6 q^4-60 q^2 r^4\Biggr]\Biggr\}
  \times \Biggl\{60 r^2 \left[q^4 r^4 \left(f_1
   r^6+3\right)+4 f_1 q^2 r^{14}+4 \left(f_1 r^{18}+r^{12}\right)+q^6\right]\Biggr\}^{-1},
    \end{aligned}
    \label{eq:arlinff}
\end{equation}
\begin{equation}
    \bar{C} (r) = \frac{q^4 r^6 \left(4 f_1 r^6+21\right)+8 f_1 q^2 r^{16}+9 q^6 r^2+12 r^{14}}{q^4 r^4
   \left(f_1 r^6+3\right)+4 f_1 q^2 r^{14}+4 \left(f_1 r^{18}+r^{12}\right)+q^6}-\frac{4
   q^2 r^2}{q^2+2 r^4}-2 r^2.
\end{equation}
\end{widetext}
 
\begin{figure}[t!]
    \centering
   \includegraphics[scale=0.65]{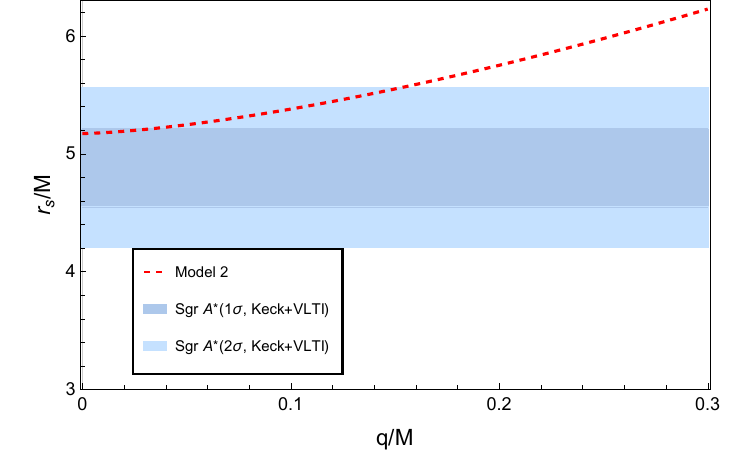}
    \caption{Same as Figure \ref{fig:16} for the effective metric on equation \eqref{eq:arlinff} with the fixed values: $f_k=2$, $a_0 =0.5$, $f_1 = 2 \times 10^{-66}$ and $\Lambda = 10^{-41}$.}
    \label{fig:19}
\end{figure}

 We calculated the shadow radius numerically and obtained the result as shown in Fig. \ref{fig:19}. Based on model 1, we have restricted the range of values of the constants $a_0$, $f_1$ and $f_k$, which is consistent with the EHT observations. For model 2 or 3, we do not need to do this again and therefore use again: $f_k=2$, $a_0 =0.5$, $f_1 = 2 \times 10^{-66}$ and $\Lambda = 10^{-41}$, in Fig. \ref{fig:19}. In this geometry, the radius of the shadow increases as we increase the charge, and we see that $r_s$ is within the $2 \sigma$ limit established by the EHT observations.

\subsection{Model 3}

\begin{figure}[t!]
    \centering
   \includegraphics[scale=0.65]{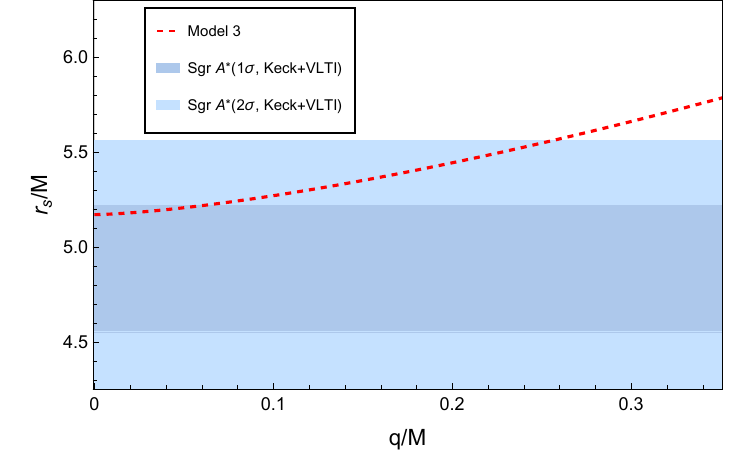}
    \caption{Same as Figure \ref{fig:16} for the effective metric on equation \eqref{eq:arsinef} with the fixed values: $f_k=2$, $a_0 =0.5$, $f_1 = 2 \times 10^{-66}$ and $\Lambda = 10^{-41}$.}
    \label{fig:20}
\end{figure}

For our last case, we substitute Eq. \eqref{eq:AsinF} in \eqref{eq:abar} and obtain

\begin{eqnarray}
	\bar{A}(r) &=& -\Bigg[2 r \left(6 M+ 2 r^3 \sin \left(\frac{q^2}{2 r^4}\right)+ \Lambda  r^3- 3 r\right)
		\nonumber\\
	&&	+ q^2
		\left(E_{\frac{3}{4}}\left(-\frac{i q^2}{2 r^4}\right)+E_{\frac{3}{4}}\left(\frac{i q^2}{2
			r^4}\right)\right)\Bigg]\Big/
	\nonumber\\
	&& \left[6 r^2 \left(f_1 r^6+\frac{q^2}{r^4}+\frac{8 r^8}{\left(q^2+2
		r^4\right)^2}-1\right) \right],
	\label{eq:arsinef}
\end{eqnarray}
\begin{equation}
	\bar{C}(r) = \frac{r^2}{-2 f_1 r^6+\frac{8}{\left(\frac{q^2}{2 r^4}+1\right)^3}+\frac{3 q^2}{r^4}-\frac{24
			r^8}{\left(q^2+2 r^4\right)^2}-1}
\end{equation}
We have again calculated the shadow radius for this geometry numerically, the behavior of $r_s$ as a function of the reduced load $q/M$ is shown in Fig. \ref{fig:20}. Increasing the load leads to an increase in the shadow and it has values within the $2 \sigma$ limit, which is therefore consistent with the observations of the EHT collaboration.

\section{Conclusion}\label{sec:conclu}

In this work, we have studied spherically symmetric and static solutions in the framework of Conformal Killing Gravity (CKG), a recently proposed modified theory of gravity coupled with nonlinear electrodynamics (NED). This new theory offers significant advances over General Relativity (GR) and allows for greater flexibility in cosmological and astrophysical modeling. One of the key aspects of this theory is that the cosmological constant emerges as an integration constant without having to be assumed beforehand, allowing a natural interpretation for dark energy. By including the NED as a matter source and considering a static, spherically symmetric line element, we have derived the relevant field equations and found a general form for the Lagrangian density, which depends on an arbitrary metric function $A(r)$ and three parameters: $f_0$ (unit of $[L]^{-1}$), $f_1$ (unit of $[L]^{-3}$) and the magnetic charge $q$ (dimensionless). We have shown that the Lagrangian densities found in the framework of CKG are more general and have a larger matter content than those leading to the same solutions in general relativity.
We have proposed 3 new solutions for black holes and for each of them we have studied the properties of the event horizons, the regularity of spacetime and the Lagrangian density aspect and also verified them using the Event Horizon Telescope (EHT) observations of Sagittarius $A^{\star}$ (Sgr $A^{\star}$).

The first model is given by the equation \eqref{eq:arexp}, where we have the parameters $M$, $q$, $\Lambda$, $a_0$ and $f_k$. If $a_0 \rightarrow 0$ the solution is simply Reissner-Nordström-Ads, and if we make $q \rightarrow 0$ we return to Schwarzschild-Ads, and therefore a generalization of these solutions is given by the new constants $a_0$ and $f_k$. This solution has up to three horizons: the event horizon, the Cauchy horizon and the cosmological horizon. We have numerically established that it is possible to reduce the number of horizons using the extremization conditions (equations \eqref{eq:con1} and \eqref{eq:con2}), and we have found at least one extreme value for each parameter of the solution. By coupling CKG with NED, we also derived the nonlinear Lagrangian density corresponding to this model, which agrees with Maxwell theory in the limit $F \rightarrow 0$ as expected. We found that increasing $f_k$ tends to increase the value of the Lagrangian. We also calculated the Kretschmann scalar and found that the spacetime generated by this solution is singular at the origin.
The second model, given by \eqref{eq:arl}, has only three parameters: $M$, $q$ and $\Lambda$; it is also a generalization of the SC-Ads solution. Again, we find three horizons and two extreme values for each parameter, as shown in Figures \ref{fig:7}, \ref{fig:9}, and \ref{fig:11}. We also derive the Lagrangian density for this model and show its behavior in the figure \ref{fig:Llinear}. We can see that changing the constant $f_1$ does not change the Lagrangian. Finally, we also show that this solution is also singular at $r =0$ by analyzing the Kretschmann scalar. The last model represented in the equation \eqref{eq:AsinF} is a solution with a metric function that has a sinusoidal part and therefore it would be possible to choose the values of the parameters ($M,q$ and $\Lambda$) to obtain a solution with many horizons. However, in order to standardize the analysis with previous models, we focused on cases where the solution has only three horizons. We found two critical charge values $q_c = 0.9958$ and $q_c = 1.0934$. We then determined the Lagrangian density generated by this solution and showed that increasing the charge slightly decreases its value, as can be seen in Figure \ref{fig:Lsin}. Furthermore, the regularity of spacetime was investigated by calculating the Kretschmann scalar, which revealed the presence of a curvature singularity in the limit $r \rightarrow 0$.

In addition, we conducted a study on the shadows produced by the 3 proposed models. We followed the procedure already known in the literature and described by \cite{Perlick:2021aok}. Since we hypothesize that these solutions are generated by NED, we also use the effective geometry formalism to compute the shadow radius. We use the data from the EHT collaboration for the supermassive object at the center of our galaxy, Sgr $A^{\star}$, and constrain the parameters of the solution. For simplicity, we assumed $M=1$ and specified the other constants in terms of mass. In Model 1, the shadow radius $r_s$ was within the $1 \sigma$ limit for the values $0<a_0<2$, $ f_k >2$, $f_1 <10^{-65}$ and $q < 0.6$. It was also within the $2 \sigma$ limit for $f_1<10^{-64}$ and $q<0.876$. For model 2, we found that the charge must be $q<0.155$ to be within the $2 \sigma$ region and $q<0.038$ to be in the $1 \sigma$ region. For Model 3, we have determined that $q < 0.256$ for $r_s$ to be in the $2 \sigma$ region and $q < 0.062$ for the $1 \sigma$ region. To summarize,
 we have adjusted the parameters and shown that the three models can match the EHT observations for Srg $A^{\star}$.

In summary, the three proposed models extend the classical black hole solutions with significant modifications, showing sensitivity to key parameters of modified gravity and nonlinear electrodynamics. In future work, we intend to regularize these solutions by adding a scalar field to the source, transforming them into black bounce solutions and study its thermodynamics, stability and quasi-normal mode. The analysis of black hole shadows also suggests the possibility of empirically testing these scenarios, especially in regimes where multiple horizons may emerge. These results point to new developments in both theoretical studies and observational frameworks, raising new questions for future investigations.


\acknowledgments

MER thanks Conselho Nacional de Desenvolvimento Cient\'ifico e Tecnol\'ogico - CNPq, Brazil, for partial financial support. This study was financed in part by the Coordena\c{c}\~{a}o de Aperfei\c{c}oamento de Pessoal de N\'{i}vel Superior - Brasil (CAPES) - Finance Code 001.
FSNL acknowledges support from the Funda\c{c}\~{a}o para a Ci\^{e}ncia e a Tecnologia (FCT) Scientific Employment Stimulus contract with reference CEECINST/00032/2018, and funding through the research grants UIDB/04434/2020, UIDP/04434/2020 and PTDC/FIS-AST/0054/2021.


\end{document}